\PassOptionsToPackage{tbtags}{amsmath}
\documentclass[aps,prx,reprint,amsmath,amssymb,groupedaddress,superscriptaddress,floatfix]{revtex4-2}
\usepackage{graphicx}
\usepackage{bm}
\usepackage{booktabs}
\usepackage{array}
\usepackage{makecell}
\usepackage{microtype}
\usepackage[colorlinks,linkcolor=blue,anchorcolor=blue,citecolor=blue,urlcolor=blue]{hyperref}

\newcommand{\ecl}{\epsilon_{\mathrm{clean}}}
\newcommand{\ep}{\epsilon_p}
\newcommand{\dn}{\Delta_{\mathrm{noise}}}
\newcommand{\dnpk}{\Delta_{\mathrm{noise}}^{\mathrm{peak}}}
\newcommand{\dnfin}{\Delta_{\mathrm{noise}}^{\mathrm{final}}}
\newcommand{\dnt}{\tilde\Delta_{\mathrm{noise}}}
\newcommand{\Sop}{S_{\mathrm{op}}}
\newcommand{\Rop}{R_{\mathrm{op}}}
\newcommand{\Eg}{E_g}
\newcommand{\tr}{\operatorname{Tr}}
\newcommand{\ket}[1]{\lvert #1\rangle}
\newcommand{\bra}[1]{\langle #1\rvert}
\newcommand{\av}[1]{\langle #1\rangle}
\newcommand{\Wrx}{W_{rx}}
\newcommand{\Wxy}{W_{xy}}
\newcommand{\Wzz}{W_{zz}}
\graphicspath{{figures/}}

\usepackage{placeins}

\begin{document}

\title{Phase Interference and Transverse Catalysts Govern the Noise Fragility of Long-Range Variational Quantum Circuits}

\author{Longhui Shen}
\affiliation{Basic Science Department, Wuchang Shouyi College, Wuhan 430064, China}

\author{Fangzhou Jin}
\affiliation{Basic Science Department, Wuchang Shouyi College, Wuhan 430064, China}

\author{Xiaoxia Yi}
\affiliation{Basic Science Department, Wuchang Shouyi College, Wuhan 430064, China}

\author{Bin Guo}
\email{binguo@whut.edu.cn}
\affiliation{Department of Physics, Wuhan University of Technology, Wuhan 430070, China}

\date{\today}

\begin{abstract}
Long-range entangling gates shorten variational state preparation, but they also accelerate operator spreading across the register, linking nonlocality to a presumed noise fragility. Here we show that the single-fault fragility $F$ of a variational circuit is governed by how fault phases interfere on the target Hamiltonian, not by how far faults spread. For Heisenberg chains we prove that two faults with identical extensive support can incur either bounded or extensive energy damage, in the exact ratio $N/2$. A phase-matched longitudinal halo remains energetically dark until ignited by transverse catalysts that break a symmetry of the remote register, namely spatially nonuniform single-qubit rotations or the pair-flip component of two-qubit $XX/YY$ gates, whereas uniform rotations and magnetization-conserving exchange gates leave it strictly dark. In noisy training with full-width long-range $ZZ$ gates, suppressing either catalyst alone leaves circuits trapped on an ignited plateau sustained by remote $XY$ couplings, whereas suppressing both eliminates the plateau, achieves minimal fragility despite larger fault operators, and halves the dephasing error. Training is driven by a race between catalyst pruning and gradient force balance, as annealing the transverse catalysts out of trapped circuits rescues every trajectory while annealing the entire longitudinal sector does not. Combined with the gate-count exposure $\lambda=pK$, a single fragility measurement accurately predicts the noise-induced energy shift across circuit architectures and noise strengths, reducing ansatz design to a trade-off between coherent expressivity and gate count.
\end{abstract}
\maketitle

\section{Introduction}\label{sec:intro}

Variational quantum algorithms prepare many-body states with parameterized circuits~\cite{Leone2024,Holmes2022,McClean2016,Kandala2017,Kokail2019} whose depth is limited by gate noise~\cite{Peruzzo2014,Cerezo2021,Bharti2022,Yan2025,Tilly2022,Preskill2018}. Long-range interactions, natively available in trapped ions, Rydberg arrays, and programmable simulators with power-law couplings $r^{-\alpha}$~\cite{Monroe2021,Browaeys2020,Qiao2024,Shen2026,Guo2025,Koyluoglu2025}, shorten the depth needed to build extensive correlations~\cite{Eldredge2017,Tran2021,Bond2025,Guo2020}. On such platforms the speed at which correlations and entanglement spread depends on the interaction exponent and can exceed the bounds of short-range systems~\cite{Richerme2014,Jurcevic2014,FossFeig2015}, and variational optimization has been demonstrated directly on long-range spin models~\cite{Pagano2020}. The same nonlocality, however, forfeits the geometric locality that confines gate errors in short-range circuits. A Pauli fault inserted early in a complete-graph circuit is conjugated by gates that touch all qubits, and standard operator-spreading measures---the Pauli support $\Sop$ and radius $\Rop$ of the propagated fault---rapidly reach the register size~\cite{Nahum2018,Keyserlingk2018,Tan2025,Chalker2026,Khemani2018}. Since a spread fault overlaps with bonds across the entire lattice, nonlocality is naturally associated with heightened noise fragility.

Whether rapid operator spreading actually translates into severe energetic degradation is, however, far from obvious. Standard spreading metrics such as $\Sop$ and $\Rop$ are Frobenius-norm quantities that count every non-identity Pauli component equally, remaining blind both to whether a propagated fault commutes with the target Hamiltonian $H$ and to the relative phases with which components on neighboring sites enter. Even for a fixed dynamics, light cones defined through different norms and operational tasks can differ qualitatively~\cite{Tran2020}, so a spreading metric is not by itself a measure of physical damage. In digital simulations and error mitigation, a propagated error need not corrupt a local observable, because only Pauli strings up to a relevant weight contribute~\cite{Granet2025,GonzalezGarcia2025} and faults outside an observable's effective light cone drop out~\cite{Eddins2024}. Complementary approaches based on noise-robust shallow shadows mitigate circuit-induced bias in the estimation of nonlocal quantum observables~\cite{Hu2025}. Variational optimization can moreover partly compensate systematic hardware errors~\cite{OMalley2016} and can be combined with active error minimization~\cite{Li2017}, and the noise resilience of variational algorithms has been traced both to the classical optimizer~\cite{Gentini2020} and to the adaptation of the optimized parameters to the noise~\cite{Fontana2021}. In variational circuits, where gate errors accumulate across the circuit volume to deform the cost landscape and induce noise-driven plateaus~\cite{Wang2021,StilckFranca2021,Cincio2021,Sharma2020,PedroGarciaPintos2025}, our companion study of nearest-neighbor Heisenberg circuits~\cite{Shen2026a} showed that confining the initial transverse generator amplitudes keeps fault operators spatially localized and avoids high-noise force-balance plateaus, whereas restricting the longitudinal angles or repeating layers is unnecessary. Long-range gates overturn the premise of that result because a fault can spread over the whole register even when every transverse angle is small, so whether operator spreading still controls the noise must be decided anew.

To separate the hardware overhead of a nonlocal circuit from the intrinsic sensitivity of its quantum state, the noise-induced energy error can be organized by a single relation. When each of the $K$ fault locations carries an error probability $p$ and removes on average a fraction $F$ of the coherent energy, independent faults~\cite{Dalton2024,Dalzell2024} compound into the noise-induced energy shift $\dn\simeq(|E_0|/|\Eg|)(1-e^{-\lambda F})$ with exposure $\lambda=pK$~\cite{Shen2026a}, which we derive in Eq.~\eqref{eq:expo}. While the hardware and gate count set $\lambda$, the circuit parameters set the single-fault fragility $F$ through the Heisenberg-picture propagation of individual Pauli faults~\cite{Rudolph2026,Angrisani2026}. The question of long-range fragility therefore reduces to a question about $F$, asking how the phases of many long-range gates interfere in the energy of an extensive fault, which gates activate a symmetry-protected dark fault, and how these mechanisms organize noisy training.

Here we answer these questions for Hamiltonian-variational circuits~\cite{Wecker2015,Wiersema2020,Park2024,Ho2019} on complete, hybrid, and nearest-neighbor interaction graphs targeting antiferromagnetic Heisenberg chains, combining exact operator identities with exact noisy training. (i) A longitudinal $ZZ$ halo is transparent to remote $ZZ$ bonds and damages remote transverse bonds only through a signed graph-Laplacian phase mismatch, so that two faults with the same extensive support $\Sop=(N+1)/2$ cost $O(1)$ or $O(N)$ energy, in the exact ratio $N/2$. (ii) A phase-matched dark halo is ignited only by transverse catalysts that break a symmetry of the remote register, namely spatially nonuniform $R_x$ rotations, whose exact finite-angle damage depends on neighboring angle differences and reduces at small angles to the ordinary graph-Laplacian form, and the pair-flip part of two-qubit $XX/YY$ gates, whereas uniform rotations and the magnetization-conserving exchange part leave the halo dark. (iii) In trained circuits that keep every long-range $ZZ$ gate at full width, either catalyst alone holds the circuit on an ignited plateau, whereas suppressing both yields the lowest fragility with larger operators, keeps fault excitations at low energy, and halves the error under dephasing noise. (iv) Noisy training is a race between catalyst pruning and force balance in which full-width remote $XY$ gates are never pruned and keep circuits trapped, whereas circuits with weak remote $XY$ gates eventually accommodate full-width rotations without pruning them; annealing the catalysts out of trapped circuits releases every seed, whereas annealing up to all longitudinal angles does not. (v) A fragility measured once predicts the noise shift across architectures and noise strengths, and protected circuits of all architectures share the same fragility, so that the architecture choice reduces to a quantitative trade-off between coherent accuracy and gate count.

The remainder of this paper is organized as follows. Section~\ref{sec:model} presents the circuit model, establishes the fragility law, and formulates two theorems, whose proofs are provided in Appendix~\ref{app:star}. Section~\ref{sec:decoupling} validates the theoretical predictions through a controlled catalyst experiment. Section~\ref{sec:dynamics} examines the training dynamics and identifies the underlying mechanisms through causal interventions. Section~\ref{sec:scaling} investigates the scaling of noise exposure and its dependence on circuit architecture. Finally, Sec.~\ref{sec:conclusion} summarizes the main findings and discusses their implications for robust variational quantum circuit design.

\section{Single-Fault Fragility Law and Exact Operator Identities}\label{sec:model}

\subsection{Circuit Model and Fragility Law}\label{sec:ansatz}

We consider $N\in\{4,6,8\}$ qubits on a periodic ring with cyclic distance $r_{ij}=\min(|i-j|,N-|i-j|)$ and target the nearest-neighbor (NN) Heisenberg antiferromagnet
\begin{equation}\label{eq:H}
H=\sum_{j=1}^{N}\left(X_jX_{j+1}+Y_jY_{j+1}+Z_jZ_{j+1}\right),
\end{equation}
with exact ground-state energy $\Eg$ ($\Eg=-14.6044$ at $N=8$). Starting from $\ket{0}^{\otimes N}$, each of $L\in\{4,8\}$ layers applies single-qubit rotations $R_x$ and $R_z$ on all sites, followed by two-qubit blocks $R_{xx}$, $R_{yy}$, and $R_{zz}$ on all $N(N-1)/2$ pairs of the complete graph, where $R_P(\theta)=\exp(-i\theta P/2)$. Within each block, pairs are ordered by distance shell, $r_{ij}=1,\dots,N/2$, and each shell is decomposed greedily into disjoint matchings. A final N\'eel decoder $D=\prod_{j\text{ odd}}R_x(\pi)$ maps $\ket{0}^{\otimes N}$ onto the reference N\'eel state, so that the variational gates act in a staggered logical frame. The complete-graph circuit contains $P=L[2N+3N(N-1)/2]$ parameters ($P=400$ at $N=8$, $L=4$). Sparse architectures are analyzed in Sec.~\ref{sec:scaling}, and long-range Heisenberg targets in Appendix~\ref{app:robust}.

Every elementary gate, including the decoder, is followed by an independent single-qubit Pauli channel on each qubit it acts upon~\cite{Wang2021,Singkanipa2025,Mele2026},
\begin{equation}\label{eq:channel}
\mathcal E(\rho)=(1-p)\rho+p\sum_{Q\in\{X,Y,Z\}}\pi_Q\,Q\rho Q,
\end{equation}
with $\sum_Q\pi_Q=1$. The circuit contains $K=L[2N+3N(N-1)]+N/2$ fault locations ($K=740$ at $N=8$, $L=4$), and we quantify the cumulative noise strength by the exposure $\lambda=pK$. With the noisy energy $E_p=\tr[H\rho_p(\bm\theta)]$ and the noiseless energy $E_0=\bra{\psi_0}H\ket{\psi_0}$ at the same parameters, the relative error $\ep=(E_p-\Eg)/|\Eg|$ splits as $\ep=\ecl+\dn$ into the coherent error $\ecl=(E_0-\Eg)/|\Eg|$ and the noise shift $\dn=(E_p-E_0)/|\Eg|$.

A single Pauli fault $Q$ at location $s$ is conjugated by the remainder of the circuit, $C_{>s}$, into the propagated operator $W_{s,Q}=C_{>s}Q_sC_{>s}^\dagger$ acting on $\ket{\psi_0}$~\cite{Rudolph2026,Angrisani2026,GonzalezGarcia2025}. Its energy damage is
\begin{equation}\label{eq:damage}
\Delta E_{s,Q}=\bra{\psi_0}W_{s,Q}^\dagger HW_{s,Q}\ket{\psi_0}-E_0,
\end{equation}
which determines the first-order susceptibility and the single-fault fragility,
\begin{align}
\chi&=\sum_{s=1}^{K}\sum_{Q}\pi_Q\,\Delta E_{s,Q},\label{eq:chi}\\
F&=\frac{\chi}{K|E_0|},\label{eq:fragility}
\end{align}
with $E_p=E_0+p\chi+O(p^2)$~\cite{Dalton2024}. The fragility is the mean fraction of the coherent energy that one fault removes, and $\overline{\Delta E}=\chi/K$ is the mean damage per fault. Suppose that each fault contracts the coherent energy by the same fraction $F$ toward its infinite-temperature value $\tr H/2^N=0$, so that a run with $n$ independent faults~\cite{Dalzell2024} retains $E_0(1-F)^n$. Averaging over the binomially distributed number of faults at the $K$ locations gives
\begin{equation}\label{eq:compound}
E_p\simeq E_0(1-pF)^K\simeq E_0\,e^{-\lambda F},
\end{equation}
and hence the fragility law
\begin{equation}\label{eq:expo}
\dn\simeq\frac{|E_0|}{|\Eg|}\left(1-e^{-\lambda F}\right),
\end{equation}
which reproduces the first-order result $E_p-E_0=p\chi$ for $\lambda F\ll1$. Because $|E_0|/|\Eg|=1-\ecl$, dividing out this coherent prefactor defines the normalized noise shift
\begin{equation}\label{eq:expo_norm}
\dnt\equiv\dn\,\frac{|\Eg|}{|E_0|}\simeq1-e^{-\lambda F},
\end{equation}
which depends on the circuit only through the product $\lambda F$ and reduces to $\dnt\simeq\lambda F$ for $\lambda F\ll1$. Since $\ep=\ecl+(1-\ecl)\dnt$, well-trained circuits with $\ecl\ll1$ obey $\ep\simeq\dnt$. The single-fault erasure fraction and its multiplicative compounding were introduced for NN circuits in Ref.~\cite{Shen2026a}, where they reproduce exact density-matrix simulations at noise rates for which the linear estimate overshoots. An exponential dependence of noisy expectation values on the expected number of faults likewise underlies exponential error extrapolation under Pauli noise~\cite{Cai2021}. Equation~\eqref{eq:expo} separates what the hardware controls, $\lambda$, from what the circuit controls, $F$; Sec.~\ref{sec:scaling} tests it across long-range and sparse architectures, interventions, and noise strengths.

\begin{figure*}[t]
\includegraphics[width=\textwidth,clip]{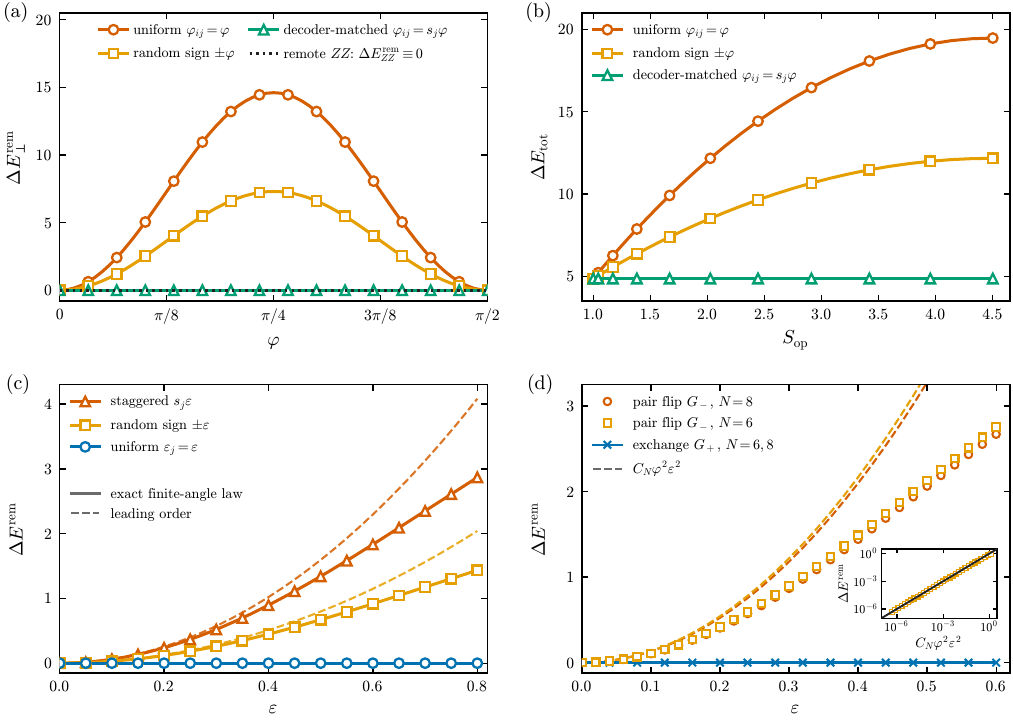}
\caption{Exact identities for a propagated $X_i$ fault on exact ground states of the Heisenberg ring. (a) Remote transverse damage $\Delta E^{\mathrm{rem}}_\perp$ behind a $ZZ$ star versus the star angle $\varphi$ at $N=8$ for uniform ($\varphi_{ij}=\varphi$), random-sign ($\pm\varphi$, exact average over all $2^{N-1}$ sign patterns), and decoder-matched ($\varphi_{ij}=s_j\varphi$) phases; symbols are exact, lines Eq.~\eqref{eq:QH}. The dotted line is the remote $ZZ$-bond damage, identically zero. (b) Total fault damage $\Delta E_{\mathrm{tot}}$, Eq.~\eqref{eq:Etot}, versus the operator support $\Sop$ at $N=8$ for $0\le\varphi\le\pi/4$; $\Sop$, $S_\perp$, $S_z$, and $\Rop$ are identical for the three phase patterns at each $\varphi$. (c) Ignition of a dark halo (decoder-matched phases $\varphi_{ij}=s_j\varphi$, $\varphi=0.4$, $N=6$) by remote rotations $R_x(\varepsilon_j)$: remote damage $\Delta E^{\mathrm{rem}}$ versus the rotation amplitude $\varepsilon$ ($\varepsilon_j=s_j\varepsilon$, $\pm\varepsilon$, or $\varepsilon$); solid curves are the exact finite-angle law Eqs.~\eqref{eq:ignite_exact} and \eqref{eq:rotation_overlap}, dashed curves its leading order Eq.~\eqref{eq:ignite}. Uniform rotations leave the halo dark. (d) Remote damage $\Delta E^{\mathrm{rem}}$ versus the angle $\varepsilon=\theta_\pm$ of a two-qubit rotation $e^{-i\varepsilon G_\pm}$ on a remote pair after a dark star ($\varphi=0.4$, physical frame): the exchange part $G_+$ leaves the halo dark, the pair-flip part $G_-$ ignites it; dashed curves are $C_N\varphi^2\varepsilon^2$, Eq.~\eqref{eq:CN}, with the independently computed coefficient $C_N$ of Eq.~\eqref{eq:CN_coef}. Inset: exact pair-flip damage $\Delta E^{\mathrm{rem}}$ versus $C_N\varphi^2\varepsilon^2$ along $\varphi=0.4s$, $\varepsilon=0.3s$, $s\in[0.025,1]$; the line has unit slope.}\label{fig:theory}
\end{figure*}

Two further diagnostics resolve the energetics of a propagated fault. On the exact ground manifold $\{\ket g\}$, its damage is the first frequency moment of the dynamical structure factor $S_W(\omega)$ of $W_{s,Q}$~\cite{Hohenberg1974},
\begin{equation}\label{eq:Eground}
\begin{split}
\Delta E_{\mathrm{ground}}&=\sum_n(E_n-\Eg)\,\overline{|\bra{n}W_{s,Q}\ket{g}|^2}\\
 &=\int d\omega\,\omega\,S_W(\omega),
\end{split}
\end{equation}
and its spectral weight in the upper half of the many-body spectrum defines the high-energy probability
\begin{equation}\label{eq:Pup}
P_\uparrow=\sum_{n:\,x_n\ge1/2}\overline{|\bra{n}W_{s,Q}\ket{g}|^2},
\end{equation}
with the normalized excitation energy $x_n=(E_n-\Eg)/(E_{\max}-\Eg)$. The geometric spreading of the fault follows from its Pauli expansion $W_{s,Q}=\sum_Pc_PP$~\cite{Nahum2018,Keyserlingk2018}. With the site weight $w_j=\sum_{P:\,P_j\neq I}|c_P|^2$, the operator support and the operator radius about the fault site $i$ are
\begin{align}
\Sop&=\sum_jw_j=S_\perp+S_z,\label{eq:support}\\
\Rop^2&=\frac{1}{\Sop}\sum_jr_{ij}^2w_j,\label{eq:radius}
\end{align}
where the transverse support $S_\perp$ and the longitudinal support $S_z$ collect the parts of $w_j$ with $P_j\in\{X,Y\}$ and $P_j=Z$, respectively. In a long-range circuit every two-qubit block can spread a fault across the register, so $\Sop$ and $\Rop$ grow rapidly. Whether this spreading raises $F$ is settled by the two exact results below.

\subsection{Theorem 1: Phase Interference of a Longitudinal Halo}\label{sec:thm1}

Consider an $X_i$ fault propagated through a star of long-range longitudinal gates $U_z=\prod_{j\neq i}\exp(-i\varphi_{ij}Z_iZ_j/2)$ in the logical frame. Because $Z_iZ_j$ anticommutes with $X_i$,
\begin{equation}\label{eq:star}
\begin{split}
W&=U_zX_iU_z^\dagger\\
 &=\prod_{j\neq i}\left(\cos\varphi_{ij}-i\sin\varphi_{ij}Z_iZ_j\right)X_i.
\end{split}
\end{equation}

Each remote site acquires a purely longitudinal weight $w_j=\sin^2\varphi_{ij}$, so that the transverse support stays at $S_\perp=1$ while the total support becomes extensive,
\begin{equation}\label{eq:star_support}
\Sop=1+\sum_{j\neq i}\sin^2\varphi_{ij}\le N.
\end{equation}

This halo commutes with every remote longitudinal bond $Z_jZ_k$ ($j,k\neq i$). As long as no transverse rotation follows the star, it is therefore exactly transparent to the longitudinal part of $H$, $\Delta E^{\mathrm{rem}}_{ZZ}\equiv0$.

On remote transverse bonds the damage is instead set by phase interference. In the logical frame the N\'eel decoder introduces the sublattice signs $s_j=(-1)^j$ and the bond signs $\sigma_{jk}=s_js_k$. Evaluating $W^\dagger HW$ in the singlet ground state gives the exact remote transverse damage
\begin{align}
\Delta E^{\mathrm{rem}}_\perp&=\frac{4|\Eg|}{3N}\,Q_{H,i},\label{eq:QH}\\
Q_{H,i}&=\sum_{\langle j,k\rangle\not\ni i}\sin^2\Delta_{jk},\label{eq:phase_mismatch}
\end{align}
where the sum runs over the NN bonds not incident on $i$ and $\Delta_{jk}=\varphi_{ij}-\sigma_{jk}\varphi_{ik}$ is the phase mismatch across the bond. At small angles the total mismatch reduces to a quadratic form,
\begin{equation}\label{eq:signed_laplacian}
Q_{H,i}\simeq\bm\varphi_i^{T}\mathcal L_H^{(\sigma)}\bm\varphi_i,
\end{equation}
where $\mathcal L_H^{(\sigma)}=S\mathcal L_HS$, with $S=\mathrm{diag}(s_j)$, is the balanced signed Laplacian~\cite{Zaslavsky1982} built from the ordinary Laplacian $\mathcal L_H$ of the remote target graph, whose kernel is the decoder-matched mode $\varphi_{ij}\propto s_j$. Energetic damage is thus a gradient functional on the target graph, whereas the support of Eq.~\eqref{eq:star_support} is a magnitude functional. For a smooth decoder-matched halo $\varphi_{ij}=s_jf(r_{ij})$ with $f\sim r^{-\beta}$, the sublattice sign cancels in $\Delta_{jk}$, which becomes a finite difference of $f$. The site weight then decays as $r^{-2\beta}$, whereas the bond mismatch obeys the derivative law
\begin{equation}\label{eq:derivative_law}
\sin^2\Delta_{jk}\simeq(\partial_rf)^2\sim r^{-2(\beta+1)},
\end{equation}
so the bond damage falls off two powers faster than the support, as Fig.~\ref{fig:derivative} confirms. The two bonds incident on $i$ are damaged exactly as by a bare $X_i$ fault, whatever the halo, so the total damage of an arbitrary star is
\begin{equation}\label{eq:Etot}
\Delta E_{\mathrm{tot}}=\frac{4|\Eg|}{3N}\bigl(2+Q_{H,i}\bigr).
\end{equation}

The contrast is sharpest at $\varphi=\pi/4$. A uniform (unif) star ($\varphi_{ij}=\varphi$) and a decoder-matched (dm) star ($\varphi_{ij}=s_j\varphi$) then share the same support $\Sop=(N+1)/2$, radius, and Pauli weights, yet give $Q_{H,i}=N-2$ and $0$, respectively. Their total damages, $O(N)$ versus $O(1)$, differ for every even $N\ge4$ by the exact factor
\begin{equation}\label{eq:ratio}
\frac{\Delta E_{\mathrm{tot}}^{\mathrm{unif}}}{\Delta E_{\mathrm{tot}}^{\mathrm{dm}}}=\frac{4|\Eg|/3}{8|\Eg|/3N}=\frac{N}{2}.
\end{equation}
Figure~\ref{fig:theory}(a,b) confirms Eqs.~\eqref{eq:star}--\eqref{eq:phase_mismatch} and \eqref{eq:Etot} against exact diagonalization (maximum residual $9\times10^{-15}$ at $N=8$; $2\times10^{-14}$ for $N=4$--$12$). In the physical frame the propagated decoder-matched star is $W=e^{-is_i\varphi A}X_i$, generated by $A=Z_iM_{\mathrm{rem}}$, where the remote magnetization $M_{\mathrm{rem}}=\sum_{j\neq i}Z_j$ commutes with every $\mathrm{U}(1)$-symmetric remote bond. Independent random signs destroy this coherence and give exactly half of the uniform remote damage. The proofs of Eqs.~\eqref{eq:QH}, \eqref{eq:derivative_law}, and \eqref{eq:Etot}, together with further exact checks, are given in Appendix~\ref{app:star}.

\subsection{Theorem 2: Catalytic Ignition by Transverse Gates}\label{sec:thm2}

A dark halo stays dark until noncommuting transverse gates convert its latent $Z$ weight into transverse defects through the nested commutator $[X_j,[Z_iZ_j,X_i]]=4Y_iY_j$. Single-qubit rotations $V=\prod_{j\neq i}R_x(\varepsilon_j)$ applied after the star rotate each remote $Z_j$ into $\cos\varepsilon_jZ_j-\sin\varepsilon_jY_j$ and thus convert a fraction $\sin^2\varepsilon_j$ of each remote longitudinal weight into transverse weight,
\begin{align}
S_\perp&=1+\sum_{j\neq i}\sin^2\varphi_{ij}\sin^2\varepsilon_j,\label{eq:sperp}\\
S_z&=\sum_{j\neq i}\sin^2\varphi_{ij}\cos^2\varepsilon_j.\label{eq:sz}
\end{align}

For a dark star, $\varphi_{ij}=s_j\varphi$, the $\mathrm{SU}(2)$ isotropy of the singlet ground state yields the remote damage---transverse plus longitudinal---in closed form,
\begin{align}
\Delta E^{\mathrm{rem}}&=\frac{4|\Eg|}{3N}\sum_{\langle j,k\rangle\not\ni i}\bigl(1-u_{jk}^2\bigr),\label{eq:ignite_exact}\\
u_{jk}&=\cos^2\varphi+\sin^2\varphi\cos(\varepsilon_j-\varepsilon_k),\label{eq:rotation_overlap}
\end{align}
which reproduces exact diagonalization to $2\times10^{-14}$ across $\varphi,\varepsilon\in[0,0.8]$, as Fig.~\ref{fig:theory}(c) shows. At small angles, $1-u_{jk}^2\simeq\varphi^2(\varepsilon_j-\varepsilon_k)^2$, and
\begin{equation}\label{eq:ignite}
\begin{split}
\Delta E^{\mathrm{rem}}&\simeq\frac{4|\Eg|}{3N}\,\varphi^2\,\bm\varepsilon^{T}\mathcal L_H\bm\varepsilon\\
 &=\frac{\varphi^2}{2}\av{[B_\varepsilon,[H_{\mathrm{rem}},B_\varepsilon]]},
\end{split}
\end{equation}
where $B_\varepsilon=\sum_{j\neq i}\varepsilon_jY_j$ in the physical frame and $H_{\mathrm{rem}}$ collects the bonds not touching $i$. Ignition is the transverse dual of Theorem~1: it is governed by the gradient of the rotation angles on the target graph, with the same prefactor $4|\Eg|/3N$ as in Eq.~\eqref{eq:QH}. Uniform rotations lie in the kernel of $\mathcal L_H$---$B_\varepsilon$ is then proportional to the total remote spin and commutes with $H_{\mathrm{rem}}$---so the halo stays dark at any angle. Staggered rotations $\varepsilon_j=s_j\varepsilon$ excite the $q=\pi$ spin-flip sector, $\Delta E^{\mathrm{rem}}\simeq c_\xi\varphi^2\varepsilon^2$ with $c_\xi=16(N-2)|\Eg|/3N$ ($39.86$ at $N=6$), and random signs give half of this value.

The same symmetry criterion classifies the two-qubit catalysts. On a remote pair $(j,k)$ the gates $R_{xx}(\theta_{xx})$ and $R_{yy}(\theta_{yy})$ commute, and in the physical frame their product factorizes exactly into an exchange and a pair-flip rotation,
\begin{equation}\label{eq:pairflip}
R_{xx}R_{yy}=e^{-i(\theta_+G_++\theta_-G_-)},
\end{equation}
with the generators $G_\pm=(X_jX_k\pm Y_jY_k)/2$ and the angles $\theta_\pm=(\theta_{xx}\pm\sigma_{jk}\theta_{yy})/2$. The exchange generator $G_+$ conserves $M_{\mathrm{rem}}$ and hence commutes with the generator $A$ of the dark star, so that $e^{-i\theta_+G_+}$ leaves the halo dark at any angle. The pair-flip generator $G_-$ changes $M_{\mathrm{rem}}$ by $\pm4$ and ignites the halo at the same mixed fourth order as a nonuniform $R_x$,
\begin{equation}\label{eq:CN}
\Delta E^{\mathrm{rem}}\simeq C_N\varphi^2\theta_-^2,
\end{equation}
whose coefficient is the remote spin stiffness of the generator $B=-i[G_-,A]$ on the bare fault state $\ket b=X_i\ket g$,
\begin{equation}\label{eq:CN_coef}
C_N=\frac{1}{2}\bra b\bigl[B,[H_{\mathrm{rem}},B]\bigr]\ket b.
\end{equation} 

Exact diagonalization with the rotation $e^{-i\varepsilon G_\pm}$ applied to a remote pair, i.e., $\varepsilon=\theta_\pm$, confirms both statements in Fig.~\ref{fig:theory}(d): the exchange damage vanishes to $4\times10^{-15}$ at $N=6$ and $8$, whereas the pair-flip damage approaches $C_N\varphi^2\varepsilon^2$ ($C_6=84.67$, $C_8=81.92$) to a relative deviation of $2\times10^{-4}$ at small angles. A single $R_{xx}(\theta)$ gate already carries $\theta_-=\theta/2$, and independently initialized $\theta_{xx}$ and $\theta_{yy}$ generically give $\theta_-\neq0$, so a full-width remote $XY$ channel is a catalyst even when the $R_x$ angles are small. The derivations of Eqs.~\eqref{eq:sperp}--\eqref{eq:CN_coef} are given in Appendix~\ref{app:star}.

Theorems~1 and 2 thus assign definite roles to the circuit generators. Long-range $R_{zz}$ gates build an extensive halo whose damage is fixed by phase interference and vanishes for the decoder-matched phase pattern. Spatially nonuniform $R_x$ rotations and the pair-flip part of remote $R_{xx}R_{yy}$ gates are catalysts that ignite it; uniform rotations and the exchange part are not. Magnetization-conserving exchange rotations are also the basic two-qubit elements of symmetry-preserving ans\"atze for the variational quantum eigensolver~\cite{Gard2020}, so the dark channel coincides with a standard symmetry-adapted gate set. These are exact statements for one gate layer acting after a dark star. In a multilayer circuit the rotation axes are scrambled by interleaved, generically nonuniform $R_z$ angles, and $XX$ and $YY$ gates act in separate blocks, so any finite-amplitude $R_x$ or remote $XX/YY$ gate is a potential catalyst and the symmetry criterion becomes a condition on gate amplitudes. Because either catalyst can ignite the halo---a logical OR---a circuit is protected by construction only when both are weak; whether a catalyst that is present actually ignites the faults of a trained circuit is decided by training, as Sec.~\ref{sec:dynamics} shows.

\section{Damage Decoupling and Dephasing Selectivity in Trained Circuits}\label{sec:decoupling}

We test these theoretical predictions in a controlled catalyst experiment summarized in Table~\ref{tab:arms}, drawing all initial angles as $\theta_g=b_g\xi_g$ with independent $\xi_g\sim U[-1,1]$. Narrow initial-angle distributions are known to mitigate vanishing gradients in deep circuits~\cite{Zhang2022}; here, by contrast, the widths are restricted selectively by generator type. In the flat-envelope design, every $R_z$ gate, every $R_{zz}$ gate at all distances, and the NN ($r=1$) $R_{xx}$ and $R_{yy}$ gates are initialized at full width $b=\pi$. Only the two potential catalysts are varied by setting the single-qubit width $\Wrx$ and the width $W_{xy}^{\mathrm{rem}}$ of the long-range ($r\ge2$) $R_{xx}$ and $R_{yy}$ gates to either $\pi$ or $0.2$ (here ``remote'' denotes long-range circuit gates with $r\ge2$, whereas in Theorems~1 and 2 it denotes target bonds not incident on the fault site). This choice of a confined width $0.2$ alongside full-width, layer-independent longitudinal angles follows the hierarchical ablation of Ref.~\cite{Shen2026a}, where bounding only the initial transverse generators to $[-0.2,0.2]$ sufficed for low-fragility training of NN circuits. All four flat-envelope arms therefore retain the full long-range longitudinal halo and differ solely in which transverse catalysts start small. To connect these controlled arms to broader distance-dependent ans\"atze, we also examine a power-law family with envelopes
\begin{align}
b^{xy}_{ij}&=\Wxy\,c(\alpha_c)\,r_{ij}^{-\alpha_c},\label{eq:envelope}\\
b^{zz}_{ij}&=\Wzz\,c(\beta_c)\,r_{ij}^{-\beta_c},\label{eq:envelope_zz}
\end{align}
in either bare ($c=1$) or row-normalized Kac~\cite{Campa2009} ($c=2/\sum_{j\neq i}r_{ij}^{-a}$) form. The six resulting power-law arms are listed in the lower half of Table~\ref{tab:arms}, and their dependence on the interaction exponents and normalization protocols is detailed in Appendix~\ref{app:robust}.

\begin{table}[t]
\caption{Initialization arms. Widths are envelope prefactors; $W_{rz}=\pi$ throughout. Upper block: flat-envelope design (all $R_{zz}$ gates and the $r=1$ $XX/YY$ gates at width $\pi$). Lower block: power-law family, Eqs.~\eqref{eq:envelope} and \eqref{eq:envelope_zz}, bare with $\alpha_c=\beta_c=0.5$ unless stated; the Kac arm uses $W=\pi(1-1/N)$ in both two-qubit channels with row normalization, the short-range arm $\alpha_c=\beta_c=3$.}\label{tab:arms}
\centering\small
\begin{tabular*}{\columnwidth}{@{\extracolsep{\fill}}lcccc@{}}
\toprule
Arm & $\Wrx$ & $W_{xy}$ ($r=1$) & $W_{xy}$ ($r\ge2$) & $\Wzz$ \\
\midrule
\multicolumn{5}{l}{Flat-envelope design} \\
Full & $\pi$ & $\pi$ & $\pi$ & $\pi$ \\
Small $R_x$ & $0.2$ & $\pi$ & $\pi$ & $\pi$ \\
Small remote $XY$ & $\pi$ & $\pi$ & $0.2$ & $\pi$ \\
Protected & $0.2$ & $\pi$ & $0.2$ & $\pi$ \\
\midrule
\multicolumn{5}{l}{Power-law family} \\
Full & $\pi$ & \multicolumn{2}{c}{$\pi$} & $\pi$ \\
Small $R_x$ & $0.2$ & \multicolumn{2}{c}{$\pi$} & $\pi$ \\
Small $XY$ & $\pi$ & \multicolumn{2}{c}{$0.2$} & $\pi$ \\
Small $R_x$+$XY$ & $0.2$ & \multicolumn{2}{c}{$0.2$} & $\pi$ \\
Kac & $\pi$ & \multicolumn{2}{c}{$\pi(1-\tfrac1N)$} & $\pi(1-\tfrac1N)$ \\
Short-range & $\pi$ & \multicolumn{2}{c}{$\pi$} & $\pi$ \\
\bottomrule
\end{tabular*}
\end{table}

\begin{figure}[htbp]
\includegraphics[width=\columnwidth,clip]{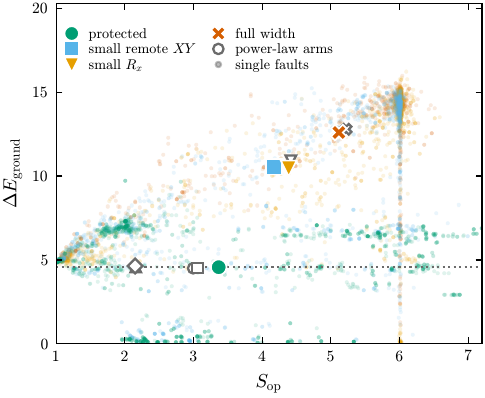}
\caption{Support--damage decoupling in the flat-envelope design ($N=8$, $L=4$, $\lambda=1.0275$, $t=600$; medians of five seeds, 114 stratified fault locations). Ground-state damage $\Delta E_{\mathrm{ground}}$ versus operator support $\Sop$ for individual faults and Pauli types (dots) and arm medians (large filled symbols) of the four arms, namely protected (small $R_x$ and small remote $XY$), small remote $XY$, small $R_x$, and full. Open gray symbols are the six power-law arms of Table~\ref{tab:fault}, and the dotted line marks $\Delta E_{\mathrm{ground}}=4.6$.}\label{fig:support}
\end{figure}

\begin{figure*}[htbp]
\includegraphics[width=\textwidth,clip]{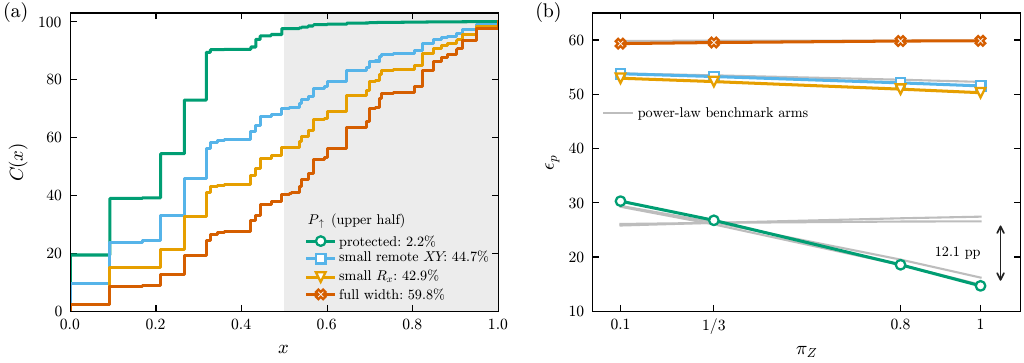}
\caption{Fault spectra and dephasing selectivity in the flat-envelope design (circuits and colors as in Fig.~\ref{fig:support}). (a) Cumulative spectral weight $C(x)$, in percent, of $W_{s,Q}\ket g$ in the eigenstates $\ket n$ of $H$ with reduced excitation energy $(E_n-\Eg)/(E_{\max}-\Eg)\le x$, averaged over faults and seeds; the shaded upper half $x\ge1/2$ defines $P_\uparrow$, whose values are listed in the legend. (b) Noisy error $\ep$, in percent, of the trained circuits replayed under Pauli channels at the training $p$, versus the $Z$ fraction $\pi_Z$ of the channel, Eq.~\eqref{eq:channel}; gray lines are the power-law arms, and the arrow marks the $12.1$~pp drop of the protected arm from depolarizing ($\pi_Z=1/3$) to pure dephasing ($\pi_Z=1$).}\label{fig:decoupling}
\end{figure*}

\begin{table*}[t]
\caption{Single-fault diagnostics of trained circuits at $N=8$, $L=4$, $\lambda=1.0275$, $t=600$ (medians over seeds, computed independently for each column; flat-envelope design five seeds, power-law family seeds 0--1). Supports, $\Delta E_{\mathrm{ground}}$, and $P_\uparrow$ are stratified averages over fault locations and Pauli types; $F_Z$ is the pure-$Z$ fragility. The column Eq.~\eqref{eq:expo} gives the predicted noise shift (for the power-law rows with $|E_0|\simeq|\Eg|$), followed by the measured noise shift, the noisy error under depolarizing and pure-dephasing noise, and the corresponding ground-state fidelities $\bra g\rho_p\ket g$. $P_\uparrow$ and the last six columns are in percent.}\label{tab:fault}
\centering\footnotesize
\setlength{\tabcolsep}{3pt}
\begin{tabular*}{\textwidth}{@{\extracolsep{\fill}}lccccccccccccc@{}}
\toprule
Arm & $\Sop$ & $S_\perp$ & $S_z$ & $\Delta E_{\mathrm{ground}}$ & $P_\uparrow$ & $F$ & $F_Z$ & Eq.~\eqref{eq:expo} & $\dn$ & $\ep^{\mathrm{depol}}$ & $\ep^{Z}$ & $\mathcal F^{\mathrm{depol}}$ & $\mathcal F^{Z}$ \\
\midrule
\multicolumn{14}{l}{Flat-envelope design} \\
Protected & 3.367 & 2.111 & 1.187 & 4.56 & 2.21 & 0.307 & 0.150 & 26.9 & 26.4 & 26.76 & 14.68 & 44.4 & 64.9 \\
Small remote $XY$ & 4.172 & 2.746 & 1.454 & 10.55 & 44.73 & 0.720 & 0.681 & 50.9 & 50.3 & 53.27 & 51.54 & 36.7 & 37.0 \\
Small $R_x$ & 4.386 & 2.904 & 1.399 & 10.45 & 42.90 & 0.714 & 0.678 & 51.8 & 51.8 & 52.34 & 50.29 & 36.6 & 38.6 \\
Full & 5.117 & 3.439 & 1.672 & 12.60 & 59.85 & 0.860 & 0.876 & 57.6 & 57.8 & 59.53 & 59.85 & 35.9 & 35.8 \\
\midrule
\multicolumn{14}{l}{Power-law family} \\
Short-range & 2.160 & 1.406 & 0.753 & 4.58 & 2.36 & 0.294 & 0.306 & 26.1 & 25.4 & 26.29 & 26.57 & 43.9 & 43.5 \\
Kac & 2.150 & 1.438 & 0.712 & 4.64 & 2.78 & 0.292 & 0.319 & 25.9 & 25.0 & 26.30 & 27.44 & 43.4 & 41.5 \\
Small $R_x$+$XY$ & 2.995 & 1.862 & 1.132 & 4.50 & 2.31 & 0.290 & 0.153 & 25.8 & 25.2 & 26.01 & 14.89 & 43.8 & 64.5 \\
Small $XY$ & 3.059 & 1.919 & 1.140 & 4.52 & 2.57 & 0.291 & 0.173 & 25.9 & 25.4 & 26.39 & 16.21 & 43.8 & 61.7 \\
Small $R_x$ & 4.418 & 2.933 & 1.485 & 10.91 & 47.16 & 0.745 & 0.721 & 53.5 & 53.2 & 53.49 & 52.28 & 36.4 & 37.9 \\
Full & 5.223 & 3.499 & 1.723 & 12.79 & 61.53 & 0.875 & 0.885 & 59.3 & 58.1 & 59.87 & 59.88 & 35.7 & 35.9 \\
\bottomrule
\end{tabular*}
\end{table*}

\begin{figure*}[t]
\includegraphics[width=\textwidth,clip]{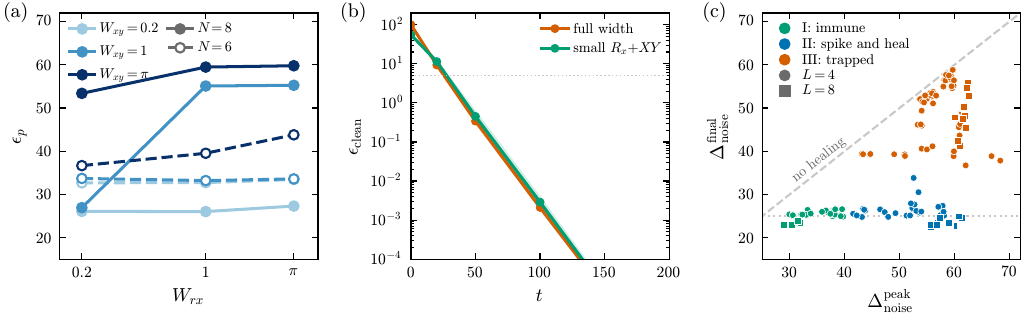}
\caption{Ignition crossover in the power-law family. (a) Noisy error $\ep$ at $t=600$ versus the single-qubit width $\Wrx$ for three transverse widths $\Wxy$ (bare, $\alpha_c=\beta_c=0.5$, $\lambda=1.0275$; medians of five seeds) at $N=8$ (filled, solid) and $N=6$ (open, dashed). (b) Noiseless control at $N=6$ and $p=0$, showing the coherent error $\ecl$ versus training update $t$ for the full and small $R_x$+$XY$ arms (medians and IQRs of 20 seeds); the dotted line marks $\ecl=5\%$. (c) Final noise shift $\dnfin$ versus transient peak $\dnpk$ of all 145 noisy $N=8$ trajectories of the power-law family ($L=4$ circles, $L=8$ squares; $\lambda=1.0275$ and fixed-$p$ runs at $\lambda=1.85$), classified as immune ($\dnpk<40\%$), spike-and-heal ($\dnpk\ge40\%$, $\dnfin<35\%$), or trapped ($\dnfin\ge35\%$); the dashed diagonal marks no healing. Errors and noise shifts are given in percent.}\label{fig:ignition}
\end{figure*}

Unless stated otherwise, circuits are trained under depolarizing noise ($\pi_Q=1/3$) at fixed exposure $\lambda=1.0275$, corresponding to $p=1.39\times10^{-3}$ at $N=8$ and $L=4$. We simulate the noisy density matrices and noiseless states exactly and optimize the parameters using exact gradients with Adam~\cite{Kingma2014} (learning rate $0.03$) for up to 1200 updates. The flat-envelope design comprises 20 trajectories across five seeds per arm; combined with the power-law family, the architectural comparisons, and the causal interventions of Secs.~\ref{sec:dynamics} and \ref{sec:scaling}, the full data set spans 1378 trajectories with five seeds per condition unless noted, from which we report seed medians and interquartile ranges (IQRs). To track the single-fault susceptibility $\chi$ efficiently, we use a stratified estimator that draws two fault locations from each stratum---defined by layer, gate type, distance shell, and decoder---and weights them by the stratum size, sampling 114 of the 740 locations at $N=8$ and $L=4$. Benchmarked against an exact evaluation over all $K=411$ fault locations across 311 snapshots at $N=6$, this stratified sample reproduces $\chi$ to within a median error of $0.6\%$ (maximum $3.4\%$) and is used for all operator and spectral diagnostics. Finally, to probe Pauli-channel selectivity without altering the learned state, we replay the trained circuit parameters unchanged under $XY$-biased ($\pi_X=\pi_Y=0.45$), $Z$-biased ($\pi_Z=0.8$), and pure-dephasing ($\pi_Z=1$) channels at the same gate error rate $p$.

As shown in Fig.~\ref{fig:support} and Table~\ref{tab:fault}, at the common comparison time $t=600$ the four arms of the flat-envelope design separate in direct agreement with the logical OR predicted by Theorem~2. The full arm is strongly ignited with $F=0.86$, and suppressing only one of the two catalysts fails to protect the circuit within this training budget. Specifically, both the small $R_x$ arm, which retains full-width remote $XY$ gates, and the small remote $XY$ arm, which retains full-width $R_x$ rotations, remain ignited at $F=0.71$--$0.72$ with ground-state damage $\Delta E_{\mathrm{ground}}\simeq10.5$ and operator support $\Sop=4.2$--$4.4$. By contrast, suppressing both catalysts simultaneously reduces the fragility to $F=0.31$ and lowers the ground-state damage to $\Delta E_{\mathrm{ground}}=4.56$. This matches the damage floor of the protected power-law arms ($4.50$--$4.64$), even though the flat-envelope protected arm carries an extensive longitudinal halo of $S_z=1.19$ and a total support of $\Sop=3.37$ that is $57\%$ larger than in the Kac and short-range arms ($\Sop\simeq2.15$). Across all protected circuits in both initialization families, $\Sop$ varies widely from $2.15$ to $3.37$ while the energetic damage remains constant to within $3\%$, as shown in Fig.~\ref{fig:support}. Energetic fragility is therefore governed by the presence of transverse catalysts rather than by operator size. Over longer training budgets the two single-catalyst arms eventually diverge in their behavior, an asymmetry that we trace to the optimization dynamics in Sec.~\ref{sec:dynamics}.

This decoupling between operator support and energetic damage is mirrored in the many-body fault spectra shown in Fig.~\ref{fig:decoupling}(a). In the protected arm, $97.8\%$ of the spectral weight of $W_{s,Q}\ket g$ remains confined to the lower half of the spectrum ($P_\uparrow=2.2\%$), matching the protected power-law arms ($P_\uparrow=2.3$--$2.8\%$). When either transverse catalyst is left at full width, however, $43$--$45\%$ of the spectral weight is pumped into the upper half of the spectrum, rising to $60\%$ in the full arm. Protected faults thus generate only low-energy excitations, whereas catalytically ignited faults populate high-energy spin-flip states. Crucially, these single-fault diagnostics quantitatively determine the macroscopic noise shift of the full multi-fault circuit. Evaluating Eq.~\eqref{eq:expo} with the measured fragility $F$ reproduces the exact noise shift $\dn$ across all 40 flat-envelope snapshots at $t=600$ and $1200$ to within $1.0$~pp (median residual $0.27$~pp), and across the power-law arms to within $1.2$~pp.

Retaining a full-width long-range longitudinal halo is not merely harmless under depolarizing noise, but actively advantageous when dephasing dominates~\cite{Bornens2025}. As Fig.~\ref{fig:decoupling}(b) shows, replaying the protected flat-envelope circuit under Pauli channels of increasing $Z$ bias reduces the relative error $\ep$ from $26.8\%$ under depolarizing noise to $14.7\%$ under pure dephasing, while boosting the ground-state fidelity from $44.4\%$ to $64.9\%$. The power-law family exhibits the same splitting among circuits that share identical depolarizing errors, as the halo-carrying small $XY$ and small $R_x$+$XY$ arms drop to $\ep=16.2\%$ and $14.9\%$ under pure dephasing, whereas the Kac and short-range arms, which suppress the longitudinal halo alongside the transverse catalysts, remain near $27.4\%$ and $26.6\%$. Ignited circuits, by contrast, remain near $\ep\simeq50$--$60\%$ regardless of the noise bias. This channel selectivity originates directly from the commutation structure of the circuit. Because a phase-flip fault $Z_i$ commutes with every $R_z$ and $R_{zz}$ gate, the longitudinal halo cannot spread it unless a noncommuting transverse gate first converts it into $X$ or $Y$ components. Consequently, early $Z_i$ faults that encounter only weak transverse catalysts nearly commute backward to the $+1$ eigenstate $\ket0^{\otimes N}$ of $Z_i$ and are largely absorbed at minimal energetic cost. In the protected arm, $Z$ faults occurring in the first two layers incur a mean damage of only $0.19$--$0.28$, whereas those occurring in the final layer or in the decoder incur $5.4$ and $4.87$, close to the bare single-site damage $8|\Eg|/3N$ given by Eq.~\eqref{eq:Etot} with $Q_{H,i}=0$. This early-layer absorption halves the pure-dephasing fragility to $F_Z=0.15$, compared with the bare value $8/3N=1/3$ retained by the Kac and short-range arms, and substituting $F_Z$ into Eq.~\eqref{eq:expo} predicts the pure-dephasing errors across all arms to within $1.5$~pp.

\section{Optimization Dynamics of the Catalyst-Pruning Race and Causal Rescue}\label{sec:dynamics}

\begin{figure*}[t]
\includegraphics[width=\textwidth,clip]{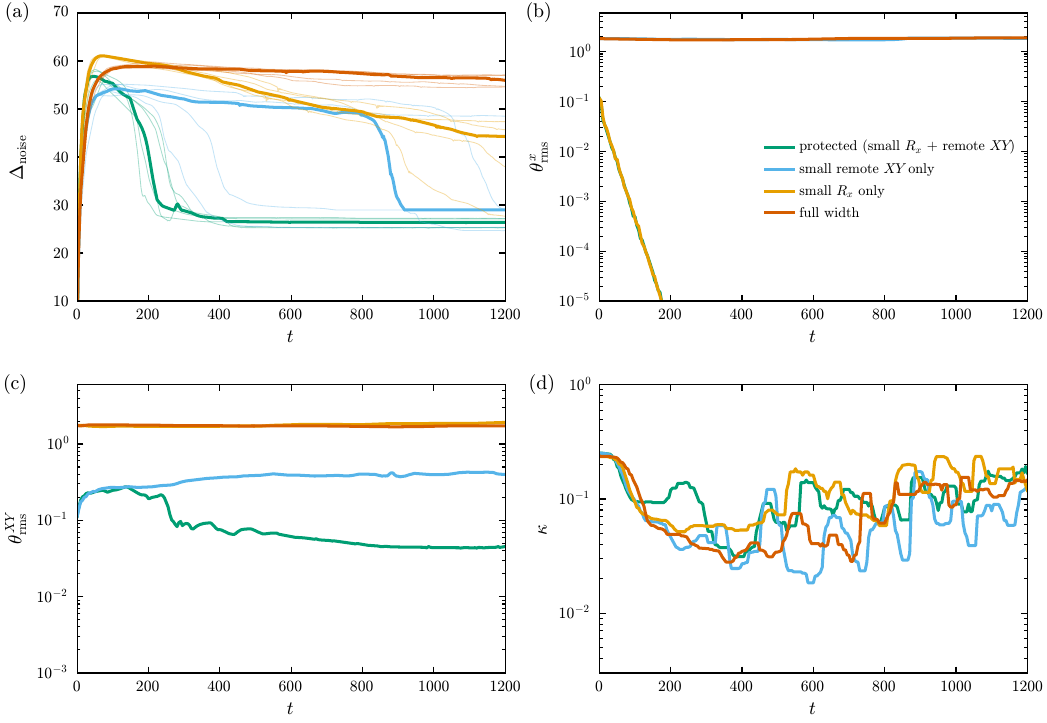}
\caption{The catalyst-pruning race in the flat-envelope design ($N=8$, $L=4$, $\lambda=1.0275$). All panels are plotted versus the training update $t$. (a) Noise shift $\dn$, in percent, showing individual seeds (thin lines) and five-seed medians (thick lines). (b) Root-mean-square $R_x$ angle $\theta^x_{\mathrm{rms}}$ and (c) root-mean-square remote $XX/YY$ angle $\theta^{XY}_{\mathrm{rms}}$ on the antipodal shell $r=4$ (medians). (d) Force-balance index $\kappa$ (median over seeds, 9-point running median). The legend in (b) applies to all panels.}\label{fig:dynamics}
\end{figure*}

Figure~\ref{fig:ignition}(a) maps how ignition in the power-law family depends on the initial catalyst widths. When the two-qubit transverse envelope is confined to $\Wxy=0.2$, increasing the single-qubit width $\Wrx$ fifteenfold leaves the noisy error nearly unchanged at $26$--$27\%$. At intermediate width $\Wxy=1$, the circuit remains protected at $\Wrx=0.2$ ($26.9\%$) but jumps by $28$~pp to $55\%$ for $\Wrx\ge1$, whereas at $\Wxy=\pi$ the two-qubit channel alone suffices to ignite the circuit, reaching $53.4\%$ already at $\Wrx=0.2$ in agreement with the pair-flip mechanism below Eq.~\eqref{eq:pairflip}. This high error in ignited circuits does not stem from a trainability bottleneck of the noiseless landscape~\cite{McClean2018,Wang2021,Larocca2025,Cerezo2025,Meyer2026,Grant2019,Larocca2022}. In noiseless controls at $p=0$, the full arm reaches the exact ground state to machine precision and is in fact the fastest to cross $\ecl\le5\%$, requiring only $26$--$28$ updates compared with $30$--$34$ updates for the small $R_x$+$XY$ arm across 20 seeds, as Fig.~\ref{fig:ignition}(b) shows. Ignition is therefore an emergent property of the noisy cost landscape~\cite{StilckFranca2021,Cincio2021}, and strategies designed to improve noiseless trainability, such as layerwise training of parameter subsets~\cite{Skolik2021}, address a different bottleneck from the catalyst selection studied here.

\begin{figure}[!tbp]
\includegraphics[width=\columnwidth,clip]{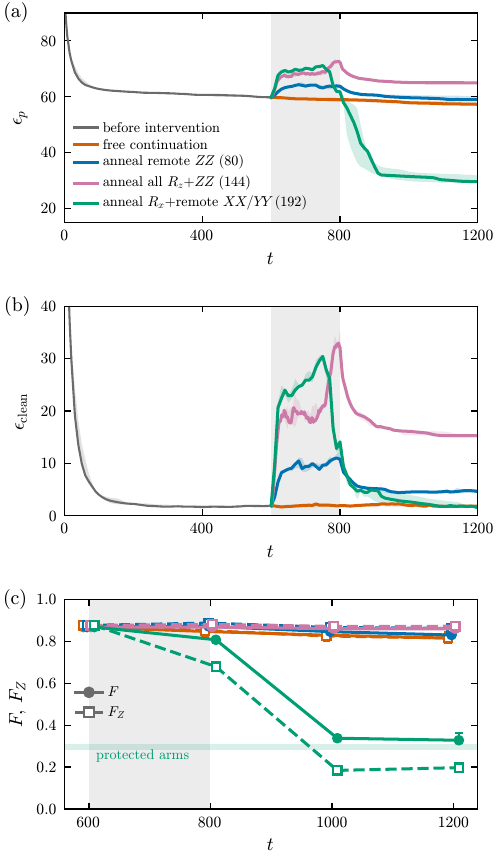}
\caption{Causal rescue of trapped circuits (power-law full arm, $N=8$, $L=4$, $\lambda=1.0275$, five seeds). From the parameters at $t=600$ (gray curve showing the common trajectory before the intervention), training continues freely or with an annealed box constraint $|\theta_g|\le|\theta_g(600)|\max[0,1-(t-600)/200]$ applied to all $R_x$ angles and the $XX/YY$ angles with $r\ge2$ (192 parameters), the $ZZ$ angles with $r\ge2$ (80), or all $R_z$ and $ZZ$ angles (144); the number of annealed parameters is given in parentheses and the shaded interval marks the annealing window. (a) Noisy error $\ep$ and (b) coherent error $\ecl$, in percent, versus training update $t$ (medians and IQRs); the legend in (a) applies to all panels. (c) Stratified fragility $F$ (filled, solid) and its pure-$Z$ component $F_Z$ (open, dashed); the horizontal band marks the protected arms of Table~\ref{tab:fault}.}\label{fig:causal}
\end{figure}

Tracking the transient peak of the noise shift $\dnpk$ alongside its final value $\dnfin$ organizes all 145 noisy $N=8$ trajectories of the power-law family into the three dynamical regimes shown in Fig.~\ref{fig:ignition}(c). Immune trajectories (33 runs) never ignite and settle directly onto the protected floor at $\dnfin\simeq25\%$. Spike-and-heal trajectories (38 runs) initially surge to $\dnpk=40$--$62\%$ before self-healing onto the same floor, whereas trapped trajectories (74 runs) heal by less than $15$~pp and remain stranded at $\dnfin=37$--$59\%$. Neither the initial angle statistics nor the height of the early noise spike determines whether an ignited trajectory heals or stalls; instead, the outcome is governed by the microscopic gate dynamics, and right at the ignition boundary with matched initial second moments, trajectories split stochastically between the healed and trapped basins independently of the range exponent, as Fig.~\ref{fig:bimodal} in Appendix~\ref{app:robust} shows.

Figure~\ref{fig:dynamics} resolves these gate dynamics catalyst by catalyst in the flat-envelope design. All four arms initially spike to $\dnpk=54$--$61\%$ as the full-width NN gates and the long-range halo scramble the input state. In the protected arm, Figs.~\ref{fig:dynamics}(b) and \ref{fig:dynamics}(c) show that Adam drives the $R_x$ angles below $10^{-5}$ within 175 updates, while the remote $XX/YY$ angles briefly grow to $0.28$ before being pruned to $0.09$ by $t=300$ and $0.05$ by $t=1200$. The noise spike in Fig.~\ref{fig:dynamics}(a) collapses within the same window, as all five seeds cross $\ep=30\%$ between updates 213 and 318 and settle at $\dnfin=25.3$--$27.2\%$. In the full arm, neither catalyst is pruned, with both remaining at root-mean-square (rms) angles near $1.8$, and no seed heals within 1200 updates ($\ep=56$--$59\%$). Between these two extremes, the single-catalyst arms reveal a pronounced dynamical asymmetry between the two catalysts. In the small $R_x$ arm, the $R_x$ angles are pruned as rapidly as in the protected arm, yet the full-width remote $XX/YY$ angles are never pruned, maintaining an rms value of $\simeq1.9$ at $r=4$ and a pair-flip angle of $|\theta_-|_{\mathrm{rms}}=1.3$--$1.5$ at $t=1200$ across every seed. Consequently, all five seeds remain trapped at $t=600$, and four of five stay above $38\%$ at $t=1200$. In the small remote $XY$ arm, where the remote $XY$ angles remain small ($0.11\to0.40$), three of five seeds still reside on the plateau at $t=600$, but four of five escape by $t=1200$ via abrupt drops at seed-dependent times ($t=370$, 634, 954, and 1187) to reach the protected floor with $F=0.29$--$0.34$. Notably, these escapes do not occur by pruning the remaining catalyst. The escaped circuits retain $O(1)$, spatially nonuniform $R_x$ angles whose mean conversion fraction $\overline{\sin^2\varepsilon_j}$ in Eq.~\eqref{eq:sperp} is $0.32$--$0.57$ at $t=1200$, comparable to the trapped seed ($0.43$) and the full arm ($0.45$--$0.64$), while the long-range $ZZ$ halo likewise retains its full width (rms angle $1.9$--$2.1$). The same behavior appears in the power-law family, where the Kac and short-range arms retain full-width $R_x$ angles (rms $\simeq1.8$ and $\overline{\sin^2\varepsilon_j}\simeq0.55$ at $t=600$) yet achieve the protected fragility $F=0.29$ listed in Table~\ref{tab:fault}. When remote $XY$ couplings are weak, a multilayer circuit can therefore absorb finite single-qubit rotations through its remaining gates and recover low fragility without pruning them. Full-width remote $XY$ gates, by contrast, drive nonlocal pair-flip transitions (Theorem~2) that the remaining gates fail to compensate in four of five seeds and that the optimizer fails to prune, forming the rigid barrier that stalls training. Suppressing the initial $R_x$ rotations alongside the remote $XY$ gates eliminates the long metastable plateau and brings every seed to the protected floor within 318 updates.

To understand why catalyst pruning stalls, we decompose the total gradient into the coherent force $g_{\mathrm{clean}}=\nabla E_0$ and the noise force $g_\delta=\nabla(E_p-E_0)$. As shown in Fig.~\ref{fig:dynamics}(d), the force-balance index $\kappa=\|g_{\mathrm{clean}}+g_\delta\|/(\|g_{\mathrm{clean}}\|+\|g_\delta\|)$ drops to $0.06$--$0.13$ across all four arms after roughly 150 updates, with the cosine of the relative angle reaching $\cos\phi=-0.98$ to $-0.99$. Because every stationary point of the noisy cost away from the noiseless ground state satisfies $g_{\mathrm{clean}}\simeq-g_\delta$, force balance characterizes all late-time plateaus and is not by itself the cause of trapping; the same first-order cancellation governs high-noise plateaus in NN circuits when updates remain confined to a manifold of constant $\ecl$~\cite{Shen2026a}. On the plateau neither gradient component vanishes ($\|g_{\mathrm{clean}}\|\simeq\|g_\delta\|\simeq0.03$--$0.05$), so a finite coherent force is locked against an opposing noise force $g_\delta\simeq\nabla\bigl[|E_0|(1-e^{-\lambda F})\bigr]$, the gradient of the noise shift in Eq.~\eqref{eq:expo}, which is driven by $\nabla F$. What distinguishes the regimes is the fragility at which this cancellation sets in. A trajectory that attains force balance after pruning the catalysts or rendering them harmless settles at $F\simeq0.3$, whereas one that reaches force balance while active catalysts persist locks in $F\simeq0.7$--$0.9$ and subsequently drifts only slowly. Noisy training is thus a race between catalyst removal and gradient force balance. The power-law family exhibits the same race, as the antipodal $XX/YY$ angles in its healing Kac arm shrink from $0.30$ to $0.06$--$0.08$ precisely while the noise spike collapses between updates 250 and 450, whereas in its trapped full arm both catalysts remain frozen near their initial values.

This race picture yields a direct causal prediction that removing the transverse catalysts from a trapped circuit should release the trajectory, whereas removing longitudinal angles should not, regardless of the perturbation size. We test this prediction in Fig.~\ref{fig:causal} on the trapped power-law full arm. Starting from the trained parameters at $t=600$ with the Adam moments restored, we continue training for 600 updates while linearly annealing a selected parameter subset to zero over 200 updates. Because the circuit architecture, fault locations, and gate error rate $p$ remain unchanged, the intervention removes gate angles rather than noise channels and allows the trajectory to depart from the constant-$\ecl$ manifold.

Annealing the transverse catalysts, comprising all $R_x$ angles and the remote $XX/YY$ angles (192 parameters), initially raises the coherent error to $\ecl\simeq30\%$ near the end of the annealing window while the optimizer reroutes the state through the remaining gates. Within about 120 updates after the window closes, the circuit heals completely. By $t=1200$ the noisy error drops to $29.6\%$ (IQR $29.2$--$31.8\%$) compared with $57.3\%$ under free continuation, representing an improvement of $22.6$--$28.2$~pp across every seed in Fig.~\ref{fig:causal}(a), while the fragility plunges from $F=0.87$ to $0.33$ ($F_Z=0.88\to0.20$) onto the protected band of Fig.~\ref{fig:causal}(c). By contrast, annealing the remote $ZZ$ angles (80 parameters) strips away the entire long-range halo but leaves the fragility at $F=0.83$ and alters the final error by only $-0.2$ to $+2.9$~pp. Even annealing all 144 longitudinal angles ($R_z$ and $ZZ$ across all distances), which inflicts a stronger coherent perturbation than the catalyst anneal with $\ecl$ peaking at $33\%$ and settling at $15\%$, leaves the circuit ignited at $F=0.86$ and raises the final noisy error to $65.0\%$ ($64.6$--$65.0\%$). Successful rescue therefore depends neither on the number of annealed parameters nor on the magnitude of the coherent perturbation; suppressing the transverse catalysts suffices to release trapped circuits, whereas removing up to the entire longitudinal sector does not. These interventions also confirm that trapping is not a failure of circuit expressibility or of the optimizer itself, since once the transverse catalysts are removed, the same Adam dynamics that stalled for 600 updates readily locates a protected solution.

\begin{figure*}[t]
\includegraphics[width=\textwidth,clip]{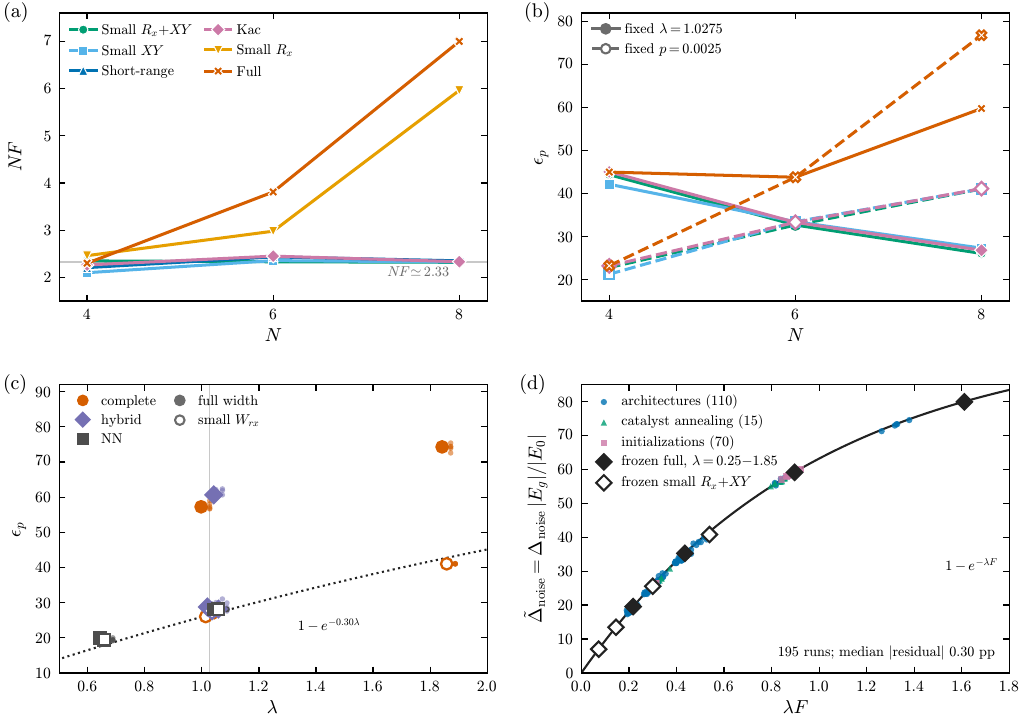}
\caption{Exposure scaling, architectures, and closure of the fragility law. (a) $NF$ versus register size $N$ for the six power-law arms at fixed $\lambda=1.0275$, $L=4$, $t=600$ (seeds 0--1); the legend in (a) applies to (a) and (b), and the gray line marks $NF\simeq2.33$. (b) Noisy error $\ep$ versus $N$ for four power-law arms at fixed $\lambda=1.0275$ (filled, solid) and at fixed $p=0.0025$ (open, dashed; $\lambda=0.45$, $1.03$, $1.85$); medians of five seeds at $t=600$. (c) Noisy error $\ep$ at $t=1200$, $N=8$, versus exposure $\lambda=pK$ for complete, hybrid (NN $XX/YY$, complete-graph $ZZ$), and NN circuits in full and small $R_x$ versions (for the complete graph also $\Wxy=0.2$), trained at fixed $p=0.0025$ ($\lambda=1.85$, $1.05$, $0.65$) and at fixed $\lambda=1.0275$ (vertical line); large symbols are medians, small dots individual seeds, and the dotted curve is the normalized fragility law of Eq.~\eqref{eq:expo_norm} with $F=0.30$, $1-e^{-0.30\lambda}$, which approximates $\ep$ for $\ecl\ll1$. (d) Normalized noise shift $\dnt$ of Eq.~\eqref{eq:expo_norm} versus $\lambda F$ for the 195 noisy trajectories of the architecture, intervention, and initialization tests (dots; $F$ measured at the end of each run) and for frozen trained parameters replayed at $\lambda=0.25$, $0.5$, $1.03$, and $1.85$ (diamonds; medians of five seeds, $F$ from all 740 fault locations); the solid curve is Eq.~\eqref{eq:expo_norm}. Errors and normalized noise shifts are given in percent.}\label{fig:scaling}
\end{figure*}

\section{Exposure Scaling and Architectural Trade-Offs}\label{sec:scaling}

The fragility law separates the two distinct ways in which a long-range circuit can degrade with system size. At fixed exposure, the mean single-fault damage of the protected arms remains size independent, with $\overline{\Delta E}=4.70\to4.34\to4.20$ for the small $R_x$+$XY$ arm as $N=4\to6\to8$ and $4.2$--$4.6$ across the remaining protected arms. Consequently, the fragility scales as $F\propto1/N$ and, as shown in Fig.~\ref{fig:scaling}(a), the product $NF$ stays near $2.33$ (within $2.10$--$2.45$) across all accessible register sizes. In ignited arms, by contrast, the single-fault damage grows rapidly with $N$ ($\overline{\Delta E}=4.60\to7.08\to12.55$ and $NF=2.30\to3.81\to7.00$ for the full arm), revealing that catalytic ignition converts a fault whose damage is bounded by a few local bonds into one whose damage scales with the system volume. Although three register sizes do not fix asymptotic scaling exponents, they clearly separate bounded from growing energetic damage. This distinction produces the opposite size trends under the two natural scaling conventions shown in Fig.~\ref{fig:scaling}(b). At fixed exposure $\lambda$, protected circuits improve with increasing $N$ ($\ep=44.4\to32.7\to26.1\%$), whereas the full arm degrades to $59.8\%$. At fixed gate error rate $p=0.0025$, however, the quadratic gate count $K\propto N^2L$ of the complete graph raises the total exposure from $\lambda=0.45$ to $1.85$, causing even the protected arms to grow in error from $22.8\%$ to $41.1\%$. Once the energetic damage per fault is bounded, the total gate count becomes the sole remaining driver of error growth.

\begin{table}[t]
\caption{Architecture comparison ($L=4$, 1200 updates, medians of five seeds). $K$ is the number of fault locations (for $N=6$/$N=8$) and $F$ the stratified fragility at $t=1200$; at $N=6$ the complete-graph fixed-$p$ and fixed-$\lambda$ runs coincide. Errors in percent.}\label{tab:arch}
\centering\footnotesize
\setlength{\tabcolsep}{2.0pt}
\begin{tabular*}{\columnwidth}{@{\extracolsep{\fill}}lccccccc@{}}
\toprule
 & & \multicolumn{3}{c}{fixed $p=0.0025$} & \multicolumn{3}{c}{fixed $\lambda=1.0275$} \\
\cmidrule(lr){3-5}\cmidrule(lr){6-8}
Architecture & $K$ & $\ep^{N=6}$ & $\ep^{N=8}$ & $F^{N=8}$ & $\ep^{N=6}$ & $\ep^{N=8}$ & $F^{N=8}$ \\
\midrule
Complete, full & 411/740 & 38.44 & 74.30 & 0.716 & 38.44 & 57.25 & 0.816 \\
Complete, small & 411/740 & 32.62 & 41.03 & 0.283 & 32.62 & 26.02 & 0.290 \\
Hybrid, full & 267/420 & 24.17 & 60.65 & 0.779 & 39.50 & 28.77 & 0.332 \\
Hybrid, small $R_x$ & 267/420 & 24.01 & 28.16 & 0.301 & 34.11 & 27.65 & 0.308 \\
NN, full & 195/260 & 18.50 & 19.85 & 0.302 & 34.05 & 28.06 & 0.281 \\
NN, small $R_x$ & 195/260 & 18.06 & 19.42 & 0.330 & 34.29 & 28.08 & 0.284 \\
\bottomrule
\end{tabular*}
\end{table}

To test this gate-count dependence directly, we prune remote two-qubit gates at the architectural level. Unlike adaptive schemes that remove or add parameters and gates during optimization~\cite{Sim2021,Bilkis2023}, this pruning fixes the connectivity, and hence the set of fault locations, before training. In hybrid circuits, the $XX$ and $YY$ gates are restricted to NN pairs while $R_{zz}$ continues to act across the complete graph; in NN circuits, all three two-qubit channels are restricted to NN pairs~\cite{Wecker2015,Wiersema2020}. Leaving all other circuit elements unchanged reduces the number of fault locations $K$ at $N=8$ from 740 in the complete graph to 420 in the hybrid architecture and 260 in the NN architecture. We train each architecture in both a full and a small $R_x$ ($\Wrx=0.2$) configuration at fixed $p=0.0025$ and at fixed $\lambda=1.0275$, alongside noiseless controls at $p=0$, and summarize the results in Table~\ref{tab:arch}.

Across all of these runs, a single quantitative relation holds. Evaluating Eq.~\eqref{eq:expo} with the fragility $F$ measured at the end of each trajectory reproduces the exact noise shift across all 195 noisy runs spanning the architectural comparisons, the causal interventions, and the phase-structured initialization tests ($\lambda=0.49$--$1.85$), yielding a median residual of $0.30$~pp and a maximum of $1.85$~pp in Fig.~\ref{fig:scaling}(d). In an even stricter test with frozen parameters, we take the trained power-law full and small $R_x$+$XY$ circuits at $t=600$, evaluate $F$ once on the noiseless state over all 740 fault locations, and replay those fixed parameters across four noise strengths from $\lambda=0.25$ to $1.85$. This single fragility measurement predicts the multi-fault noise shift at every noise level to within $0.93$~pp (median residual $0.07$~pp) over a regime in which $\dn$ spans from $7\%$ to $79\%$ (diamonds). Moreover, all protected circuits converge to nearly the same fragility, $F=0.28$--$0.33$ at $N=8$ and $0.39$--$0.42$ at $N=6$, regardless of whether long-range gates are present. Crucially, this universal low fragility is actively selected by the noisy cost function rather than built into the circuit topology, since the same hybrid and NN architectures trained without noise ($p=0$) achieve $\ecl=0.2$--$1.3\%$ ($10.9\%$ for the full hybrid circuit) but end up $1.4$--$3$ times more fragile ($F=0.45$--$0.47$ for NN and $0.87$ for hybrid at $N=8$), consistent with the pruning race of Sec.~\ref{sec:dynamics}.

Comparing the three architectures also isolates the two structural ingredients of Theorem~2. Full NN circuits retain large transverse gates but lack a long-range longitudinal halo and therefore never ignite ($F\le0.32$ across all seeds). Hybrid circuits, which add a complete-graph $ZZ$ halo to the same NN transverse gates, ignite in three of five seeds at fixed $p$ ($F\simeq0.8$, $\ep\simeq61\%$) and in one of five seeds at fixed $\lambda$, shifting the five-seed median in Table~\ref{tab:arch} from the ignited branch ($60.7\%$) to the healed branch ($28.8\%$). Weakening the single-qubit $R_x$ catalyst protects the hybrid architecture across every seed. By eliminating the remote $XY$ catalysts by construction and confining $R_x$, the protected hybrid circuit retains the entire long-range longitudinal halo at minimal fragility while cutting the number of fault locations by $43\%$ relative to the complete graph.

Because all protected circuits share a common fragility, Eq.~\eqref{eq:expo} reduces the choice of circuit architecture to a quantitative trade-off between coherent expressivity and gate count. At fixed gate error rate $p$, the gate count dominates the total error shown in Fig.~\ref{fig:scaling}(c), where at $N=8$ the NN circuits reach $\ep=19.4$--$19.9\%$, the small $R_x$ hybrid circuit reaches $28.2\%$, and the protected complete-graph arm reaches $41.0\%$ at exposures $\lambda=0.65$, $1.05$, and $1.85$, respectively. All of these low-fragility medians, together with those trained at fixed $\lambda$, lie within $2.2$~pp of the normalized fragility law of Eq.~\eqref{eq:expo_norm} with $F=0.30$, $1-e^{-0.30\lambda}$, as expected from $\ep\simeq\dnt$ for $\ecl\ll1$. At fixed exposure $\lambda$, by contrast, the architectural ordering reverses, and the protected complete-graph circuit performs best ($26.0\%$ versus $27.7$--$28.1\%$) because its long-range gates achieve a lower coherent error at equal noise budget ($\ecl=0.7\%$ versus $1.6$--$4.4\%$). Once the transverse catalysts are suppressed, long-range $ZZ$ gates thus do not inflate the intrinsic fragility $F$ and instead purchase coherent accuracy per unit exposure, paying off at fixed $p$ whenever the coherent error they eliminate exceeds the gate-count noise they introduce. With $F\simeq0.30$, Eq.~\eqref{eq:expo_norm} implies that adding $\Delta K$ fault locations at gate error $p$ raises the normalized noise shift by $e^{-\lambda F}-e^{-(\lambda+p\Delta K)F}$, which predicts a $9.3$~pp increase when moving from the NN ($K=260$) to the hybrid ($K=420$) circuit at $p=0.0025$, closely matching the measured $8.6$~pp difference between the median $\dnt$ of the small $R_x$ hybrid and the NN circuits. For the NN target Hamiltonian studied here, the NN ansatz already reaches $\ecl\le0.3\%$, leaving no room for coherent gains to offset the extra gate noise; long-range target Hamiltonians and depth-limited regimes, where short-range circuits cannot reach low $\ecl$, are where this balance can tip in favor of nonlocal gates. Across all of these regimes, the trained fragility is governed by catalyst suppression and gate count rather than by static phase patterns imposed on the initial $ZZ$ angles, as demonstrated in Appendix~\ref{app:phaseinit}.

\section{Discussion and Conclusions}\label{sec:conclusion}

Separating the noise-induced energy shift of a variational circuit into the hardware exposure $\lambda=pK$ and the single-fault fragility $F$ isolates the gate-count overhead of a nonlocal ansatz from the intrinsic sensitivity of the quantum state it prepares. Within the Heisenberg circuits studied here, we have shown that $F$ is governed by phase interference and symmetry breaking on the target Hamiltonian rather than by the spatial extent of propagated fault operators. At the single-fault level, operator support weights all Pauli components equally, whereas energetic damage is a phase-difference functional on the target graph. Because a longitudinal $ZZ$ halo commutes with remote longitudinal bonds and its remote damage vanishes on the kernel of the signed graph Laplacian, two faults with the same extensive support can incur either bounded $O(1)$ or extensive $O(N)$ energy damage, in the exact ratio $N/2$ of Eq.~\eqref{eq:ratio}. A phase-matched halo remains energetically dark until subsequent transverse gates break a symmetry of the remote register. Its damage is driven by the gradient of the single-qubit rotation angles on the target graph or by the pair-flip component of remote two-qubit $XX$ and $YY$ gates, whereas spatially uniform rotations and magnetization-conserving exchange gates leave the halo strictly dark. This protection is intrinsic to the circuit and differs from symmetry verification, which detects symmetry-breaking errors through additional measurements and postselection~\cite{BonetMonroig2018}.

In multilayer circuits, these microscopic rules translate into a condition on whether the transverse catalysts survive noisy optimization. When all nonlocal $ZZ$ gates are initialized at full width, suppressing only one of the two catalysts leaves the circuit stranded on a high-error plateau. Full-width remote $XY$ couplings act as a rigid barrier that gradient descent fails to prune, whereas full-width single-qubit rotations are eventually absorbed by the remaining circuit gates once remote $XY$ couplings are weak. Suppressing both catalysts at initialization eliminates this long-lived metastable plateau and brings the circuit rapidly to the protected fragility floor. In this protected regime, the circuit tolerates larger fault operators, confines fault excitations predominantly to the lower half of the many-body spectrum, and cuts the energy error in half under pure dephasing because early phase-flip faults nearly commute backward through the longitudinal halo to stabilize the initial state. Trajectory-resolved dynamics and causal interventions confirm that noisy training is a race between catalyst removal and gradient force balance. Annealing the transverse catalysts out of trapped circuits releases every trajectory onto the protected floor, whereas annealing remote or even all longitudinal angles never restores trainability.

These results establish three practical principles for designing nonlocal variational circuits. First, initialization should suppress both transverse catalysts while retaining the longitudinal halo. Keeping single-qubit $R_x$ rotations and the pair-flip component of remote $XX$ and $YY$ gates weak on top of full-width nonlocal $ZZ$ gates prevents persistent catalytic trapping and rapidly heals the initial noise spike, whereas suppressing only one catalyst risks prolonged plateau trapping and stripping away the longitudinal halo forfeits dephasing protection without lowering depolarizing fragility. Second, circuit connectivity should be budgeted by weighing coherent energy gain against gate-count exposure. Because protected circuits on complete, hybrid, and nearest-neighbor graphs all converge to nearly the same fragility, measuring $F$ in Eq.~\eqref{eq:expo} accurately predicts the noise shift across architectures and noise levels, pricing every additional fault location. At fixed gate error rate $p$, deploying nonlocal $ZZ$ gates is advantageous whenever the coherent error they eliminate exceeds this exposure penalty, which makes the hybrid architecture without remote $XY$ gates a natural compromise that preserves the longitudinal halo at a reduced gate count. Third, reaching this low-fragility manifold requires training directly under the noisy cost function, as neither noiseless optimization nor the static decoder-matched phase initialization of Appendix~\ref{app:phaseinit} guides deep circuits into the protected basin.

Our numerical demonstrations rely on exact density-matrix simulations of up to $N=8$ qubits under local Pauli channels and exact gradients, and our architectural comparisons focus primarily on a nearest-neighbor target Hamiltonian where a short-range ansatz already achieves high coherent accuracy. Determining how the balance between coherent expressivity and gate exposure tips for nonlocal target Hamiltonians in larger registers or under strict depth constraints, where symmetry and locality already limit the expressivity of shallow ans\"atze~\cite{Bravyi2020}, is the most direct question for future study. Combining the single-fault fragility framework and catalyst diagnostics with tensor-network~\cite{Cichy2026} or Pauli-propagation methods~\cite{Fontana2025,Schuster2025,Rudolph2026,Angrisani2026,Shrikhande2026}, alongside finite-shot gradient estimation~\cite{Sweke2020,Kaminishi2026,Schuld2019} and spatially correlated hardware noise~\cite{Oda2026}, will allow these mechanisms to be tested at the scales of current quantum processors.

\appendix
\section{Proofs of the Exact Fault Identities and the Continuum Limit}\label{app:star}

This appendix proves Theorems~1 and 2 of Sec.~\ref{sec:model}, reports the exact numerical checks, and derives the continuum derivative law.

\begin{figure}[!htb]
\includegraphics[width=\columnwidth,clip]{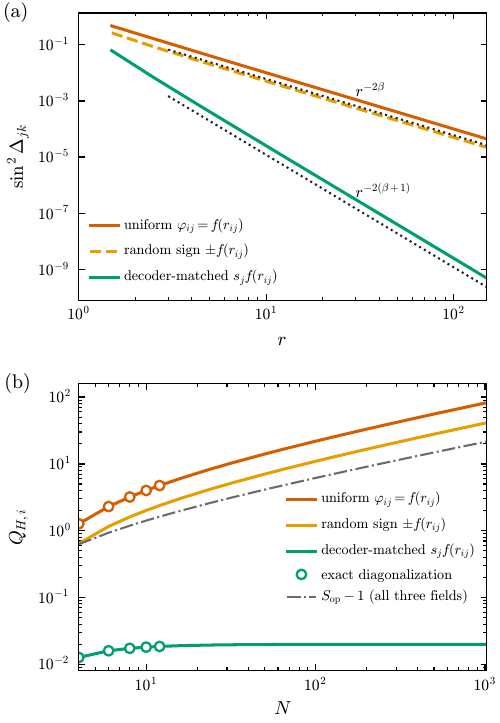}
\caption{Derivative law for decoder-matched phase fields. (a) Remote bond mismatch $\sin^2\Delta_{jk}$, with $\Delta_{jk}=\varphi_{ij}-\sigma_{jk}\varphi_{ik}$, proportional to the bond damage by Eq.~\eqref{eq:QH}, versus the distance $r$ of the bond from the fault, for $f(r)=0.5\,r^{-1}$ ($\beta=1$) at $N=400$, where the three fields have identical $\Sop$; dotted lines are the support scaling $r^{-2\beta}$ and the derivative law $r^{-2(\beta+1)}$ of Eq.~\eqref{eq:derivative_law}. (b) Total remote phase mismatch $Q_{H,i}$ versus register size $N$ for $\beta=0.25$; lines are Eq.~\eqref{eq:phase_mismatch}, circles exact diagonalization, and the dash-dotted line the common support $\Sop-1$ of all three fields, which grows as $N^{1/2}$.}\label{fig:derivative}
\end{figure}

\begin{figure*}[t]
\includegraphics[width=\textwidth,clip]{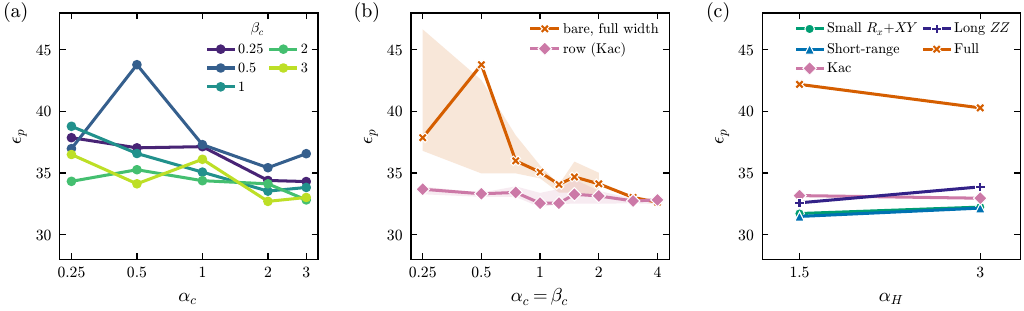}
\caption{Robustness of the power-law family at $N=6$, $L=4$, $p=0.0025$, $t=600$; all panels show the noisy error $\ep$, in percent, on a common vertical scale. (a) $\ep$ of bare full circuits versus the transverse exponent $\alpha_c$ for five longitudinal exponents $\beta_c$ (seed medians). (b) Bare full versus row-normalized (Kac) circuits as a function of $\alpha_c=\beta_c$; shading shows the IQR over five seeds. (c) $\ep$ versus the exponent $\alpha_H$ of long-range Heisenberg targets for five initialization arms.}\label{fig:robust}
\end{figure*}

\emph{Remote bonds in Theorem~1.} The N\'eel decoder $D=\prod_{j\,\mathrm{odd}}R_x(\pi)$ maps $Y_j\to-Y_j$ and $Z_j\to-Z_j$ on the odd sites, so that a physical transverse bond $X_jX_k+Y_jY_k$ becomes, in the logical frame,
\begin{equation}\label{eq:app_hjk}
h_{jk}=J_{jk}^\perp\bigl(X_jX_k+\sigma_{jk}Y_jY_k\bigr),
\end{equation}
with $\sigma_{jk}=s_js_k$, $s_j=(-1)^j$, and $J^\perp_{jk}=1$ on NN bonds. For a remote bond ($j,k\neq i$), only the star factors at $m=j,k$ fail to commute with $h_{jk}$. Since $h_{jk}$ connects only basis states with $Z_k=-\sigma_{jk}Z_j$, conjugation by $W=U_zX_iU_z^\dagger$ gives
\begin{equation}\label{eq:app_bond}
\begin{split}
W^\dagger h_{jk}W&=\cos(2\Delta_{jk})\,h_{jk}\\
 &\quad+\sin(2\Delta_{jk})\,\mathcal J_{jk},
\end{split}
\end{equation}
with $\Delta_{jk}=\varphi_{ij}-\sigma_{jk}\varphi_{ik}$ and the $Z_i$-dressed spin current $\mathcal J_{jk}=ih_{jk}Z_iZ_j$. Being Hermitian and purely imaginary in the $Z$ basis, $\mathcal J_{jk}$ has vanishing expectation value in any state with real $Z$-basis amplitudes, such as the nondegenerate singlet ground state of Eq.~\eqref{eq:H}. With $\cos(2\Delta_{jk})-1=-2\sin^2\Delta_{jk}$ and $|\av{h_{jk}}|=2|\Eg|/3N$ on every NN bond,
\begin{equation}\label{eq:app_QH}
\begin{split}
\Delta E^{\mathrm{rem}}_\perp&=2\sum_{\langle j,k\rangle\not\ni i}|\av{h_{jk}}|\sin^2\Delta_{jk}\\
 &=\frac{4|\Eg|}{3N}\,Q_{H,i},
\end{split}
\end{equation}
which is Eq.~\eqref{eq:QH}. Remote $ZZ$ bonds commute with the diagonal star and with $X_i$, which gives the longitudinal transparency.

\emph{Incident bonds and Eq.~\eqref{eq:Etot}.} For the two bonds $(i,k)$ incident on the fault site, $U_z$ commutes with $Z_iZ_k$, so
\begin{equation}\label{eq:app_incident_zz}
W^\dagger Z_iZ_kW=X_iZ_iZ_kX_i=-Z_iZ_k,
\end{equation}
exactly as for a bare $X_i$ fault. On the transverse part, conjugation by $W$ multiplies the flip-flop operator $X_iX_k-Y_iY_k$ produced by $X_i$ by a $Z$-diagonal phase operator $\Phi(\{Z_m\})$. Because $X_iX_k-Y_iY_k$ changes the total magnetization $M=\sum_mZ_m$ by $\pm4$ while $\Phi$ conserves it, the singlet ($M=0$) ground state gives, for any halo phases including $\Phi=I$,
\begin{equation}\label{eq:app_incident_perp}
\bra g(X_iX_k-Y_iY_k)\,\Phi\ket g=0.
\end{equation}

The two incident bonds therefore contribute $2\times4|\Eg|/3N$ whatever the star, which together with Eq.~\eqref{eq:app_QH} gives Eq.~\eqref{eq:Etot}. Exact diagonalization of the periodic ring for $N=4$--$12$, for uniform and decoder-matched stars at $\varphi=\pi/8$ and $\pi/4$ and for four sets of independent random phases $\varphi_{ij}\in[-\pi,\pi)$ per size, confirms Eq.~\eqref{eq:Etot} to $1.9\times10^{-14}$. At $\varphi=\pi/4$ both stars have $\Sop=(N+1)/2$ and the damage ratio equals the value $N/2$ of Eq.~\eqref{eq:ratio} to machine precision, e.g., $28.733$ versus $4.789$ at $N=12$.

\begin{figure*}[htbp]
\includegraphics[width=\textwidth,clip]{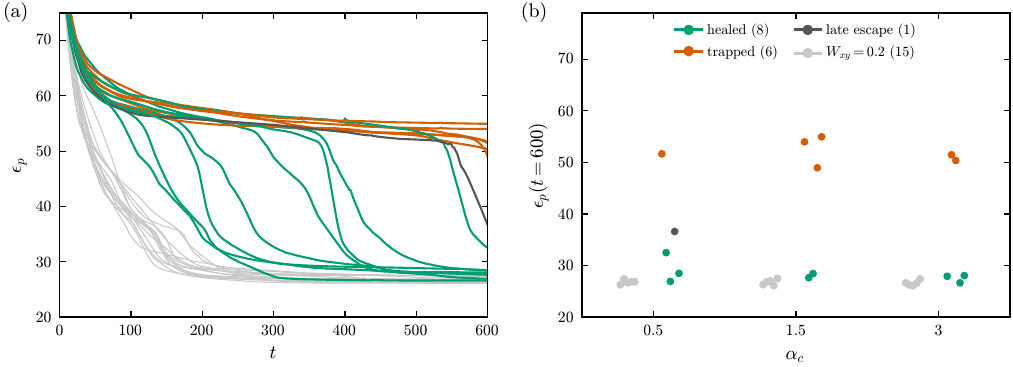}
\caption{Bimodal splitting at the ignition boundary ($N=8$, $L=4$, $\lambda=1.0275$, $q_2$ normalization). (a) Noisy error $\ep$, in percent, versus training update $t$ for the 15 trajectories at $\Wxy=1$ ($Q_2=2/3$, $\Wrx=\pi$, $\alpha_c\in\{0.5,1.5,3\}$, five seeds each), colored by outcome; gray lines are the 15 trajectories at $\Wxy=0.2$ ($Q_2=0.027$). (b) Noisy error $\ep(t=600)$, in percent, versus the range exponent $\alpha_c$; the legend gives the number of trajectories in each class and applies to both panels.}\label{fig:bimodal}
\end{figure*}

\emph{Single-qubit catalysts in Theorem~2.} Applying $V=\prod_{j\neq i}R_x(\varepsilon_j)$ after the star rotates each remote $Z_j$ into $\cos\varepsilon_jZ_j-\sin\varepsilon_jY_j$, which gives the exact weight conversion of Eqs.~\eqref{eq:sperp} and \eqref{eq:sz}; exact diagonalization confirms both to $8\times10^{-15}$. For a dark star, $\varphi_{ij}=s_j\varphi$, the decoder signs of halo and rotation cancel in the physical frame ($s_j^2=1$). Within each eigenspace $Z_i=\tau=\pm1$, the remote factors of the propagated fault become single-site unitaries
\begin{equation}\label{eq:app_wj}
w_j(\tau)=\exp\bigl[-i\tau s_i\varphi\,(\hat{\bm n}_j\cdot\bm\sigma_j)\bigr],
\end{equation}
i.e., rotations by $2\varphi$ about $\hat{\bm n}_j=(0,-\sin\varepsilon_j,\cos\varepsilon_j)$ with $3\times3$ rotation matrices $R_j(\tau)$. Factors on other sites and the fault $X_i$ commute with a remote bond and cancel, so that within each sector
\begin{equation}\label{eq:app_rotbond}
W^\dagger(\bm\sigma_j\cdot\bm\sigma_k)W=\sum_{\alpha\beta}\bigl[R_j^{T}R_k\bigr]_{\alpha\beta}\,\sigma_j^\alpha\sigma_k^\beta .
\end{equation}

In the singlet ground state, correlators of an odd number of spin operators vanish, so the projector onto $Z_i=\tau$ contributes only its identity part, and two-site correlators on NN bonds are isotropic,
\begin{equation}\label{eq:app_isotropy}
\av{\sigma_j^\alpha\sigma_k^\beta}=\frac{\delta_{\alpha\beta}}{3}\av{\bm\sigma_j\cdot\bm\sigma_k}=-\delta_{\alpha\beta}\,\frac{|\Eg|}{3N}.
\end{equation}

The bond energy is thus $-(|\Eg|/3N)\,\tr(R_j^TR_k)$. The relative rotation $w_j^\dagger w_k$ has the scalar part
\begin{equation}\label{eq:app_ujk}
u_{jk}=\tfrac12\tr\bigl(w_j^\dagger w_k\bigr)=\cos^2\varphi+\sin^2\varphi\,\hat{\bm n}_j\cdot\hat{\bm n}_k,
\end{equation}
independent of $\tau$, which is Eq.~\eqref{eq:rotation_overlap} with $\hat{\bm n}_j\cdot\hat{\bm n}_k=\cos(\varepsilon_j-\varepsilon_k)$, and its rotation matrix has trace $\tr(R_j^TR_k)=4u_{jk}^2-1$. Subtracting the unperturbed bond energy $-|\Eg|/N$ gives the damage per remote bond
\begin{equation}\label{eq:app_bond_damage}
\begin{split}
\delta E_{jk}&=-\frac{|\Eg|}{3N}\,\tr\bigl(R_j^TR_k\bigr)+\frac{|\Eg|}{N}\\
 &=\frac{4|\Eg|}{3N}\bigl(1-u_{jk}^2\bigr);
\end{split}
\end{equation}
summing over remote bonds yields Eq.~\eqref{eq:ignite_exact}. At small angles, $1-u_{jk}^2\simeq\varphi^2(\varepsilon_j-\varepsilon_k)^2$ gives the Laplacian form of Eq.~\eqref{eq:ignite}. The same leading term follows from expanding the halo generator, rotated by $V$, to second order in $\varphi$, which yields the double commutator in the second line of Eq.~\eqref{eq:ignite}. This remote spin stiffness differs from the first frequency moment of the transverse dynamical structure factor~\cite{Hohenberg1974} only by the two bonds incident on $i$. Both forms reproduce the exact coefficient $c_\xi=39.86$ at $N=6$.

\emph{Two-qubit catalysts in Theorem~2.} In the logical frame the gates on a remote pair are $R_{xx}(\theta_{xx})=e^{-i\theta_{xx}X_jX_k/2}$ and $R_{yy}(\theta_{yy})=e^{-i\theta_{yy}Y_jY_k/2}$; they commute because $X_jX_k$ and $Y_jY_k$ commute, and the decoder maps $Y_jY_k\to\sigma_{jk}Y_jY_k$. Writing $X_jX_k=G_++G_-$ and $Y_jY_k=G_+-G_-$ with $G_\pm=\tfrac12(X_jX_k\pm Y_jY_k)$, which commute with each other, gives Eq.~\eqref{eq:pairflip}. In the physical frame the propagated dark star is $W=e^{-is_i\varphi A}X_i$ with $A=Z_iM_{\mathrm{rem}}$; below, the sign $s_i$ is absorbed into $\varphi$. In terms of spin ladder operators,
\begin{align}
G_+&=\sigma_j^+\sigma_k^-+\sigma_j^-\sigma_k^+,\label{eq:app_ladder}\\[3pt]
G_-&=\sigma_j^+\sigma_k^++\sigma_j^-\sigma_k^-.\label{eq:app_ladder_minus}
\end{align}

$G_+$ conserves $Z_j+Z_k$, so $[G_+,A]=0$, and a rotation $e^{-i\varepsilon G_+}$ before and after the star cancels against itself on the remote register,
\begin{equation}\label{eq:app_exchange}
e^{-i\varepsilon G_+}e^{-i\varphi A}X_i\,e^{i\varepsilon G_+}=e^{-i\varphi A}X_i .
\end{equation}

The exchange part therefore leaves the dark halo exactly dark. $G_-$ changes $Z_j+Z_k$ by $\pm4$ and does not commute with $A$. Expanding the conjugated fault to second order in $\varphi$ and $\varepsilon$ gives the mixed fourth-order damage $C_N\varphi^2\varepsilon^2$ of Eq.~\eqref{eq:CN}, with the coefficient $C_N$ of Eq.~\eqref{eq:CN_coef}, which we evaluate independently of the finite-angle result. For the pair $(j,k)=(1,3)$ with $i=0$, exact diagonalization gives $C_6=84.67$ and $C_8=81.92$; the exact damage approaches $C_N\varphi^2\varepsilon^2$ to a relative deviation of $2\times10^{-4}$ along $\varphi=0.4s$, $\varepsilon=0.3s$ as $s\to0$, and is $0.869$ at $N=8$, $\varphi=0.4$, $\varepsilon=0.3$, whereas the exchange damage stays below $4\times10^{-15}$ for all angles, as Fig.~\ref{fig:theory}(d) shows.

\emph{Continuum derivative law.} Figure~\ref{fig:derivative} evaluates Eq.~\eqref{eq:phase_mismatch} for three phase fields of identical magnitude $|\varphi_{ij}|=f(r_{ij})=0.5\,r_{ij}^{-\beta}$, which therefore have identical $\Sop$, $S_\perp$, $S_z$, $\Rop$, and second moments. Because $\sigma_{jk}=-1$ on the NN chain, the phases of uniform and random-sign fields add across a bond, and the bond mismatch decays as $f^2\sim r^{-2\beta}$. For the decoder-matched field $\varphi_{ij}=s_jf(r_{ij})$ the sublattice sign cancels and the mismatch becomes a finite difference,
\begin{equation}\label{eq:app_derivative}
\Delta_{jk}=s_j\bigl[f(r_{ij})-f(r_{ik})\bigr]\simeq\pm s_j\,\partial_rf,
\end{equation}
which gives the derivative law of Eq.~\eqref{eq:derivative_law}, so the bond damage decays two powers faster than the support, as Fig.~\ref{fig:derivative}(a) shows; fits for $10\le r\le150$ give the exponents $-2.00$ and $-4.00$ at $\beta=1$, $-0.99$ and $-3.00$ at $\beta=0.5$, and $-4.00$ and $-6.00$ at $\beta=2$. Even for the strongly nonlocal envelope $\beta=0.25$, whose support grows as $\Sop-1\propto N^{1/2}$, the remote mismatch of the decoder-matched halo saturates at $Q_{H,i}\simeq0.02$, whereas uniform and random-sign halos of identical support grow without bound in Fig.~\ref{fig:derivative}(b). Exact diagonalization reproduces the summed mismatch for $N\le12$ to $10^{-14}$.

\section{Robustness and Ignition Boundary Statistics of the Power-Law Family}\label{app:robust}

\emph{Exponents and normalization.} For bare full circuits at $N=6$ the error is maximal in the nonlocal regime, $\ep=43.8\%$ at $\alpha_c=\beta_c=0.5$, and lies between $32.7\%$ and $38.8\%$ elsewhere, as Fig.~\ref{fig:robust}(a) shows. Row (Kac) normalization removes the peak: along $\alpha_c=\beta_c\in[0.25,4]$ it keeps $\ep=32.6$--$33.7\%$, and Fig.~\ref{fig:robust}(b) shows that the two protocols converge for $\alpha_c\ge3$. A third normalization, $q_2$, with
\begin{equation}\label{eq:app_q2}
c=\Bigl[2\Big/\sum_{j\neq i}r_{ij}^{-2a}\Bigr]^{1/2},
\end{equation}
fixes the expected second moment of the initial angles per site independently of the exponent,
\begin{equation}\label{eq:app_Q2}
Q_2=\sum_{j\neq i}\mathbb E\,\theta_{ij}^2=\frac{2W^2}{3}.
\end{equation} 

With 20 seeds, the bare full circuit reaches $40.1\%$ ($36.5\%$) at 600 (1200) updates, whereas $q_2$ or row normalization of the transverse channel alone gives $36.8\%$ ($35.4\%$) and $36.1\%$ ($35.3\%$), and $q_2$ normalization of both channels $38.4\%$ ($36.5\%$). Normalizing the transverse channel is thus at least as effective as normalizing both, in line with the benign role of the longitudinal channel.

\emph{Long-range targets.} For the long-range Heisenberg targets
\begin{align}
H_{\alpha_H}&=\sum_{i<j}J_{ij}\bigl(X_iX_j+Y_iY_j+Z_iZ_j\bigr),\label{eq:app_HalphaH}\\[3pt]
J_{ij}&=\frac{2\,r_{ij}^{-\alpha_H}}{\sum_{k\neq i}r_{ik}^{-\alpha_H}},\label{eq:app_Jij}
\end{align}
for which every row sum equals that of the NN chain, the catalyst-suppressing circuits in Fig.~\ref{fig:robust}(c) again outperform the full circuit: at $\alpha_H=1.5$ the small $R_x$+$XY$, Kac, short-range, and long $ZZ$ ($\alpha_c=3$, $\beta_c=0.5$) circuits reach $31.7$, $33.2$, $31.5$, and $32.6\%$, against $42.2\%$ for the full circuit; at $\alpha_H=3$, $32.2$, $33.0$, $32.2$, and $33.9\%$ against $40.3\%$.

\begin{figure}[!t]
\includegraphics[width=\columnwidth,clip]{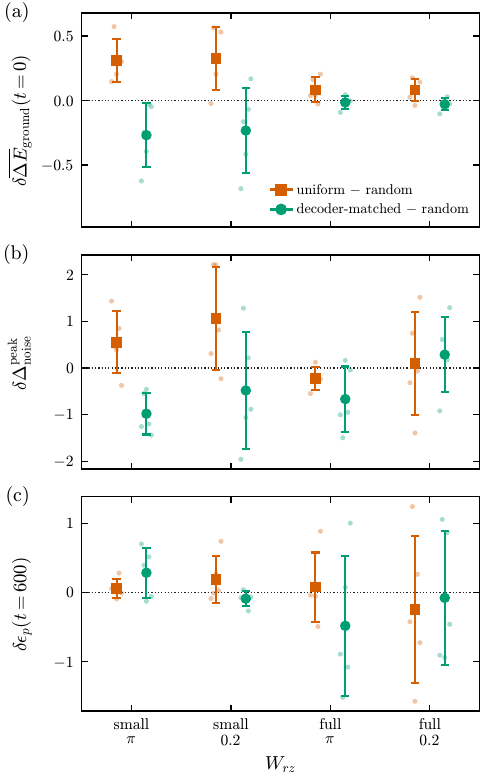}
\caption{Phase-structured initialization ($N=8$, $L=4$, $\lambda=1.0275$, five paired seeds per group). Shifts $\delta$ with respect to random $ZZ$ signs for uniform signs (squares) and decoder-matched signs $s_is_j$ (circles): (a) stratified mean ground-state damage, $\delta\overline{\Delta E}_{\mathrm{ground}}(t=0)$; (b) transient peak of the noise shift, $\delta\dnpk$; (c) noisy error at $t=600$, $\delta\ep(t=600)$; the shifts in (b) and (c) are in percentage points. The horizontal axes list the four groups: the small $R_x$+$XY$ (small) and full (full) arms, each with $R_z$ width $W_{rz}=\pi$ or $0.2$ (second line). Large symbols are means with standard deviations; faint dots are individual seeds; the legend in (a) applies to all panels.}\label{fig:phaseinit}
\end{figure}

\emph{Transverse versus longitudinal range.} At $N=8$, $L=8$, an arm with long-range transverse and short-range longitudinal envelopes (long $XY$: $\alpha_c=0.5$, $\beta_c=3$) stays at $48.5\%$ after 1200 updates, whereas the opposite arm (long $ZZ$: $\alpha_c=3$, $\beta_c=0.5$) heals by $19.5$~pp to $41.7\%$; at $L=4$ the ordering is the same. At $N=6$ ($r_{\max}=3$) the asymmetry is not resolved: the 20-seed medians show the $N=8$ ordering at $p=0.001$ (long $XY$ $20.1\%$ versus long $ZZ$ $17.5\%$) but are indistinguishable at $p=0.0025$ ($34.8\%$ versus $34.1\%$), because at this size all 25 bare full conditions of Fig.~\ref{fig:robust}(a,b) complete their spike-and-heal cycle within the budget (peak $51.9$--$60.9\%$, healing $19.6$--$26.2$~pp).

\emph{Boundary statistics.} At matched $Q_2$ the range exponent has no resolvable effect, and Fig.~\ref{fig:bimodal} shows that the outcome at the ignition boundary is stochastic. At $\Wxy=0.2$ all 15 trajectories ($\alpha_c\in\{0.5,1.5,3\}$) collapse onto $\ep=26.1$--$27.5\%$. At $\Wxy=1$ all 15 first rise to a common peak, $\dnpk=46.5$--$54.1\%$; eight then self-heal to $\ep=26.6$--$32.5\%$, with seven of them crossing $30\%$ between updates 251 and 504, six remain trapped at $48.9$--$55.0\%$, and one escapes only after $t\simeq550$. Escape is an abrupt drop at a seed-dependent time, characteristic of a long-lived plateau, and the healed fractions ($3/5$, $2/5$, $3/5$ for $\alpha_c=0.5$, $1.5$, $3$) do not depend on the exponent.

\section{Phase-Structured Initialization}\label{app:phaseinit}

Theorem~1 suggests exploiting the phase mismatch at initialization~\cite{Volkoff2021}. In the power-law full and small $R_x$+$XY$ arms ($N=8$, $L=4$, $\lambda=1.0275$, five paired seeds) we keep the magnitudes of the initial $ZZ$ angles and all other angles and assign the $ZZ$ signs randomly (the default), uniformly, or decoder-matched, $\mathrm{sgn}\,\theta^{zz}_{ij}=s_is_j$, with $W_{rz}=\pi$ or $0.2$ to vary the scrambling by $R_z$ rotations; all choices share $Q_2$ and $\sum\theta_{zz}^2=201.2$. All 60 configurations were trained for 600 updates (the ten random-sign baselines at $W_{rz}=\pi$ are reused from the main data set).

At $t=0$ the damage in Fig.~\ref{fig:phaseinit}(a) follows the ordering predicted by Eq.~\eqref{eq:QH}: in the small arm, $\Delta E_{\mathrm{ground}}$ is $0.31$--$0.33$ higher for uniform and $0.23$--$0.27$ lower for decoder-matched signs than for random signs (decoder-matched lower in 9 of 10 paired seeds), a $3\%$ effect; in the full arm, where the catalysts dominate, the shifts are below $1\%$. The decoder-matched initialization lowers the transient noise peak of the small arm by $0.5$--$1.0$~pp in Fig.~\ref{fig:phaseinit}(b), advances the first passage of $\ep\le30\%$ by $17$--$22$ updates, and reaches $t=600$ with about half the $ZZ$ angle budget ($\sum\theta_{zz}^2=172$--$184$ against $317$--$328$), but the noisy error at $t=600$ in Fig.~\ref{fig:phaseinit}(c) is unchanged within $\pm0.5$~pp in all four groups, and all three sign choices converge to the same arm-specific fragility ($F\simeq0.29$ for the small groups and $F\simeq0.87$ for the full groups). The initial phase structure remains visible in the trained $ZZ$ angles but not in the damage: in deep trained circuits it is a weak lever compared with the catalysts and the gate count.

\FloatBarrier
\section*{Acknowledgments}
This work is supported by the National Natural Science Foundation of China (Grant No. 11975175).

The authors have no conflicts to disclose.

\section*{Data Availability}
There are no publicly available research data or software supporting this manuscript. Requests for further information or data should be sent to the authors.

\bibliography{reference}

\end{document}